\documentclass[aps,
 reprint,
 superscriptaddress,
 showpacs,
 amsmath,amssymb,
 aps,
]{revtex4-1}

\usepackage[normalem]{ulem}
\usepackage{graphicx}
\usepackage[normalem]{ulem}
\usepackage{dcolumn}
\usepackage{bm}
\usepackage{hyperref}

\usepackage{epstopdf}
\usepackage{todonotes}
\usepackage{siunitx}
\usepackage{units}

\newcommand{\p}{\hphantom{\ensuremath{-}}}
\newcommand{\pd}{{\phantom{\dagger}}}

\newcommand{\kf}{\ensuremath{k_\mathrm{F}}}
\newcommand{\tk}{\ensuremath{T_\mathrm{K}}}
\newcommand{\xik}{\ensuremath{\xi_\mathrm{K}}}
\newcommand{\expect}[1]{\ensuremath{\langle #1 \rangle}}
\newcommand{\tr}[1]{\ensuremath{\textrm{Tr}\left[#1\right]}}

\begin{document}


\title{Equilibrium and real-time properties of the spin correlation function in the 
two-impurity Kondo model}

\author{Benedikt Lechtenberg}
\affiliation{Department of Physics, Kyoto University, Kyoto 606-8502, Japan}
\author{Frithjof B. Anders}
\affiliation{Lehrstuhl f\"ur Theoretische Physik II, Technische Universit\"at Dortmund, 44221 Dortmund,Germany}

\date{\today}

\begin{abstract}

We investigate the equilibrium and real-time properties of the spin correlation function $\expect{\vec{S}_1\vec{S}_2}$ in the 
two-impurity Kondo model
for different distances $R$ between the two-impurity spins.
It is shown that the competition between the Ruderman-Kittel-Kasuya-Yosida (RKKY) interaction 
and the Kondo effect governs the amplitude of $\expect{\vec{S}_1\vec{S}_2}$.
For distances $R$ exceeding the Kondo length scale, the Kondo effect also has a profound effect on the sign of the correlation function.
For ferromagnetic Heisenberg couplings $J$ between the impurities and the conduction band, 
the Kondo effect is absent and the correlation function only decays for distances beyond a certain length scale introduced by finite temperature.
The real-time dynamics after a sudden quench of the system reveals
that correlations propagate through the conduction band with Fermi velocity.
We identify two distinct timescales for the long time behavior which reflects that for small $J$ the system is driven by the RKKY interaction
while for large $J$ the Kondo effect dominates.
Interestingly, we find that at certain distances a one-dimensional
dispersion obeying
$\epsilon(k)=\epsilon(-k)$ may lead to a local parity conservation
of the impurities such that $\expect{\vec{S}_1\vec{S}_2}$
becomes a conserved quantity for long times and does not decay to its equilibrium value.

\end{abstract}


\maketitle

\section{Introduction}
\label{sec:Introduction}
Quantum impurity systems are 
promising candidates for the realization of solid-state-based quantum bits \cite{lit:Loss_DiVincenzo1998,lit:Loss_DiVincenzo1999,lit:Trauzettel2007,lit:Greilich2006,lit:Glazov2013}.
The perspective of combining traditional electronics with novel spintronics devices leads to an intense research of controlling and switching magnetic properties of such systems.
Magnetic properties of adatoms on surfaces 
\cite{lit:RevModPhys.76.323,lit:Spintronic_anisotropy_Misiorny2013,lit:Graphene_spintronics_Han2014,lit:Metals_on_silicon_Johll2014,lit:Defect_graphene_Yazyev2007,lit:Bork_Kroha2011}
or magnetic molecules 
\cite{lit:lechtenberg_2016_dimer,lit:spinfilter2010,lit:spintronic_review_Bogani2008,lit:Chemistry_Review_spintronics_Sanvito2011,lit:Review_Naber2007,lit:Organic_Dediu2009, lit:spin_diffusion_Drew2009,lit:Spinelli2015}
might serve as the smallest building blocks for such devices.

From a theoretical perspective, the two-impurity Kondo model (TIKM) \cite{lit:Jones88,lit:Jones88,lit:Jones1989,lit:Fye1989,lit:Fye1994,lit:Affleck95}
constitutes an important but simple system  which embodies the competition of interactions between two 
localized magnetic moments with those between the impurities and the conduction band.
The TIKM has been viewed as a paradigm model for the formation of two different singlet phases separated by a quantum critical point (QCP):
a Ruderman-Kittel-Kasuya-Yosida (RKKY) \cite{lit:RudermanKittel1954,lit:Kasuya1956,lit:Yosida1957} interaction induced singlet and a Kondo singlet \cite{Doniach77}.
This quantum critical point investigated by Jones and Varma (see Refs. \cite{lit:Jones88,lit:Jones1989,lit:Sakai1992_I}), however,
turned out to be unstable against particle-hole (PH) symmetry breaking \cite{lit:Affleck95}. 
The two different singlet phases are adiabatically connected
by a continuous variation of the scattering phase.
This led to the conclusion that for finite distances between the impurities no QCP exists, and the original finding is 
just a consequence of unphysical approximations \cite{lit:Fye1994}
which is generically replaced by a crossover regime \cite{lit:Silva1995}. 
Only recently, it has been shown \cite{lit:lechtenbergAnders17} that for certain dispersions and distances between the impurities the TIKM exhibits a QCP between two orthogonal ground states with different degeneracy.

In this paper, we examine the equilibrium as well as non-equilibrium properties of the spin-correlation function $\langle \vec{S}_1\vec{S}_2\rangle(R)$ for different distances $R$ between both impurity spins
using the numerical renormalization group (NRG) \cite{lit:WilsonNRG,lit:BullaReview} and 
its extension to the non-equilibrium dynamics, the time-dependent NRG (TD-NRG) \cite{lit:Anders05,lit:Anders06}.
Previously, the spatial dependence of the equilibrium properties has been mainly studied
using a simplified density of states (DOS) \cite{lit:Jones87,lit:Jones88,lit:Jones1989}  
that suppresses the antiferromagnetic (AFM) correlations \cite{lit:Affleck95}. 
In this paper, we include the full energy dependency of the even- and odd-parity
conduction-band DOSs
that properly encode the ferromagnetic (FM) as well as the antiferromagnetic 
contributions to the RKKY interaction. This approach
generates the correct RKKY interaction and does not require adding an artificial spin-spin interaction
to account for this term \cite{lit:Jones87,lit:Jones88,lit:Jones1989}.

In order to set the stage for the investigation of the non-equilibrium quench dynamics, we
present results  for the impurity spin-correlation function
$\langle \vec{S}_1\vec{S}_2\rangle(R)$. For an isotropic 
dispersion in one dimension, we find that the amplitude of $\langle \vec{S}_1\vec{S}_2\rangle(R)$ is completely
governed by the ratio between the distance of the impurities and the Kondo length scale $R/\xik$.
$\xik=v_\mathrm{F}/\tk$ is often referred to as the size of the Kondo screening cloud 
where $v_\mathrm{F}$ denotes the Fermi velocity of the metallic host and $\tk$ denotes the Kondo temperature.
For small distances $R<\xik$ and vanishing temperature, 
step like oscillations between ferromagnetic and antiferromagnetic correlations 
can be observed for $\langle \vec{S}_1\vec{S}_2\rangle(R)$ due to the RKKY interaction.
Interestingly, at large distances $R\geq\xik$ 
the ferromagnetic correlations vanish 
and only small antiferromagnetic correlations between the impurities
are found. These weak antiferromagnetic correlations are related to the PH symmetry breaking 
in the two parity channels and vanish for $R\to \infty$.

For a ferromagnetic coupling between the impurities and the conduction band, 
the Kondo effect is absent, and  a constant amplitude for the correlations is observed
at zero temperature even for $R\to \infty$. A finite
temperature introduces a new length scale beyond which correlations are exponentially suppressed.

The time dynamics of the correlation function $\langle \vec{S}_1\vec{S}_2\rangle(R,t)$ is
examined after a quench in the coupling strength between the impurities and the conduction band,
starting from initially decoupled impurities.
Experimentally, such quenches can be realized with strong laser light \cite{lit:Takasan2017}.
We have identified two distinct timescales characterizing the long-time behavior:
The RKKY interaction drives the dynamics for small Kondo coupling
whereas  a timescale $\propto 1/\sqrt{\tk}$ indicates that the physics is dominated by the Kondo effect
at large Kondo coupling.

The correlation function approaches its equilibrium value in the steady state
for most distances. For special $R$, however,
it remains almost constant although the RKKY interaction reaches a ferromagnetic maximum for those distances.
Focusing on a dispersion of a  inversion symmetric one-dimensional (1D) lattice,
parity conduction-band states decouple from the impurities at low temperatures, thus enforcing a local impurity parity conservation
such that $\langle \vec{S}_1\vec{S}_2\rangle$ becomes a conserved quantity for long times.

We combined the time-dependent correlation functions for different but fixed distances
into a two-dimensional (2D) spatial-temporal picture of the real-time dynamics. It allows for
better visualization of the the propagation of correlations. Starting from a distance   
around $\kf R/\pi=0.5$ a ferromagnetic correlation emerges which afterwards propagates with the 
Fermi velocity $v_\mathrm{F}$, defining a light cone \cite{lit:LiebRobinsonBound72,lit:lechtenbergAnders14}, 
through such a fictitious two-impurity Kondo system with variable impurity distance $R$.

\section{Model and Methods}
\label{sec:ModelMethod}

\subsection{Mapping the model onto an effective two-band model}\label{sec:ModelMethod:Mapping}

While Wilsons original NRG approach \cite{lit:WilsonNRG} was only designed to solve the thermodynamics of one localized impurity, 
the NRG was later successfully extended by Jones and Varma (see Refs. \cite{lit:Jones87,lit:Jones88,lit:Jones1989,lit:Affleck95, lit:Borda07,lit:lechtenbergAnders14,lit:lechtenberg_2016_dimer})
to two impurities separated by a distance $R$.
For this purpose the conduction band is divided into two bands, one with even-parity and one with odd-parity symmetry, the effective DOSs of which incorporated the spatial extension.
In the following, we briefly summarize this procedure for the TIKM.

The Hamiltonian of the TIKM can be separated into three parts $H=H_\mathrm{c} + H_\mathrm{int} + H_\mathrm{d}$.
$H_c$ contains the conduction-band $H_c=\sum_{\vec{k},\sigma}\epsilon_{\vec{k}}c^{\dagger}_{\vec{k},\sigma}c^{\p}_{\vec{k},\sigma}$
where $c^\dagger_{\vec{k},\sigma}$ creates an electron with spin $\sigma$ and momentum $\vec{k}$.
The interaction between the conduction band and the impurities is given by
\begin{align}
	H_\mathrm{int}=&  J \left( \vec{S}_1 \vec{s}_c(\vec{R}_1) + \vec{S}_2 \vec{s}_c(\vec{R}_2) \right), \label{eq:app:Mapping_TIKM:general_H}
\end{align}
where the impurity $\vec{S}_i$ located at position $\vec{R}_i$ is coupled via  
the effective Heisenberg coupling $J$ to the unit-cell volume averaged conduction electron spin $\vec{s}_c(\vec{r})=V_u \vec{s}(\vec{r})$.
Here, $\vec{s}(\vec{r})$ is the conduction band spin density operator expanded in planar waves
\begin{align}
	\vec{s}(\vec{r}) =& \frac{1}{2} \frac{1}{NV_u}\sum_{\sigma\sigma'}\sum_{\vec{k}\vec{k}'} 
		      c^\dagger_{\vec{k}\sigma} [\vec{ {{ \sigma}} }]_{\sigma\sigma'}  c_{\vec{k}'\sigma'} e^{i(\vec{k}'-\vec{k})\vec{r}},
\end{align}
with $N$ being the number of unit cells in the volume $V$, $V_u=V/N$ the volume of such a unit cell, $\vec{k}$ a momentum vector and $\vec{ {{ \sigma}} }$ a vector of the Pauli matrices.
In the following, we set the origin of the coordinate system in the middle of the two impurities such that $\vec{R}_1=\vec{R}/2$ and $\vec{R}_2=-\vec{R}/2$.
\\
$H_\mathrm{D}$ comprises all contribution acting only on the impurities
\begin{align}
	H_\mathrm{d}=& K \vec{S}_1\vec{S}_2, \label{eq:H_Imp}
\end{align}
with the direct Heisenberg interaction $K$ between two-impurity spins.
Unless stated otherwise, we use $K=0$ throughout this paper.

Instead, the correlations between the two-impurity spins are caused by the indirect Heisenberg interaction $K_\mathrm{RKKY} \propto J^2$ 
which is mediated by the conduction-band electrons \cite{lit:RudermanKittel1954,lit:Kasuya1956,lit:Yosida1957}.

Exploiting the symmetry
\cite{lit:JayaprakashKrischnamurtyWilkins1981,lit:Jones87,lit:Jones88,lit:Jones1989,lit:Affleck95, lit:Borda07,lit:lechtenbergAnders14,lit:lechtenberg_2016_dimer},
the conduction electron band is mapped onto
the two distance and energy dependent
orthogonal even-parity ($e$) and odd-parity ($o$) parity eigenstate field operators
\begin{align}
c_{\sigma,e/o } (\epsilon)
=& \sum_{\vec{k}} \delta(\epsilon-\epsilon_{\vec{k}}) c_{\vec{k},\sigma}  
			  \frac{\left( e^{+ i\vec{k}\vec{R}/2} \pm e^{- i\vec{k}\vec{R}/2} \right)}{N_{e/o}(\epsilon, \vec{R})\sqrt{N\rho_{c}(\epsilon)}}.
\end{align}
Here $\rho_c(\epsilon)$ is the DOS of the original conduction band and the dimensionless normalization functions are defined as
\begin{subequations}
\label{eq:NormFactor:ab}
\begin{align}
N^2_{e}(\epsilon,\vec{R})= &\frac{4}{N\rho_{c}(\epsilon)} \sum_{\vec{k}} \delta(\epsilon-\epsilon_{\vec{k}}) \cos^2\left(\frac{\vec{k}\vec{R}}{2}\right) \\
N^2_{o}(\epsilon,\vec{R})=& \frac{4}{N\rho_{c}(\epsilon)} \sum_{\vec{k}} \delta(\epsilon-\epsilon_{\vec{k}}) \sin^2\left(\frac{\vec{k} \vec{R}}{2}\right) \label{eq:NormFactor}
\end{align}
\end{subequations}
such that $c_{\sigma,e/o}(\epsilon)$ fulfill the standard anticommutator relation 
$\{c^{\phantom{\dagger}}_{\sigma,p}(\epsilon),c^\dagger_{\sigma',p'}(\epsilon')\}=\delta_{\sigma,\sigma'}\delta_{p,p'}\delta(\epsilon-\epsilon')$.
With these even- and odd-parity conduction band states the interaction part of the Hamiltonian reads
\begin{align}
 \label{eq:TIKM:H_I}
 H_\mathrm{int}=& \frac{ J}{8} \int \int \; d\epsilon \; d\epsilon' \sqrt{\rho_c(\epsilon) \rho_c(\epsilon')} \sum_{\sigma\sigma'} \vec{\sigma}_{\sigma\sigma'} \\
	   \times \left[ (\vec{S}_1 \right. & \left.  +\vec{S}_2) \sum_p \left( N_p(\epsilon,R)N_p(\epsilon',R) c^\dagger_{\sigma,p}(\epsilon)c^{\pd}_{\sigma',p} (\epsilon')\right) \right. 
	   \nonumber \\
	  \left.+ (\vec{S}_1  \right. & \left.  -\vec{S}_2) N_e(\epsilon,R)N_o(\epsilon',R) \left( c^\dagger_{ \sigma,e}(\epsilon)c^{\pd}_{\sigma',o} (\epsilon')
	      + \mathrm{h.c.} \right) \right]  \, .
	       \nonumber
\end{align}
It is important to note that due to the energy dependent factors $N_p(\epsilon,R)$, the model will generally be particle-hole asymmetric 
even if the original conduction band and, therefore, the original DOS $\rho_c(\epsilon)$ are particle-hole symmetric.
For $N_e(\epsilon,R)\neq N_o(\epsilon,R)$ this asymmetry will generate potential scattering terms that are different for the even and odd conduction bands 
and lead to the destruction of the Jones and Varma QCP (see Refs. \cite{lit:Affleck95,lit:Eickhoff2018}).

\begin{figure}[t]
	\includegraphics[width=0.5\textwidth]{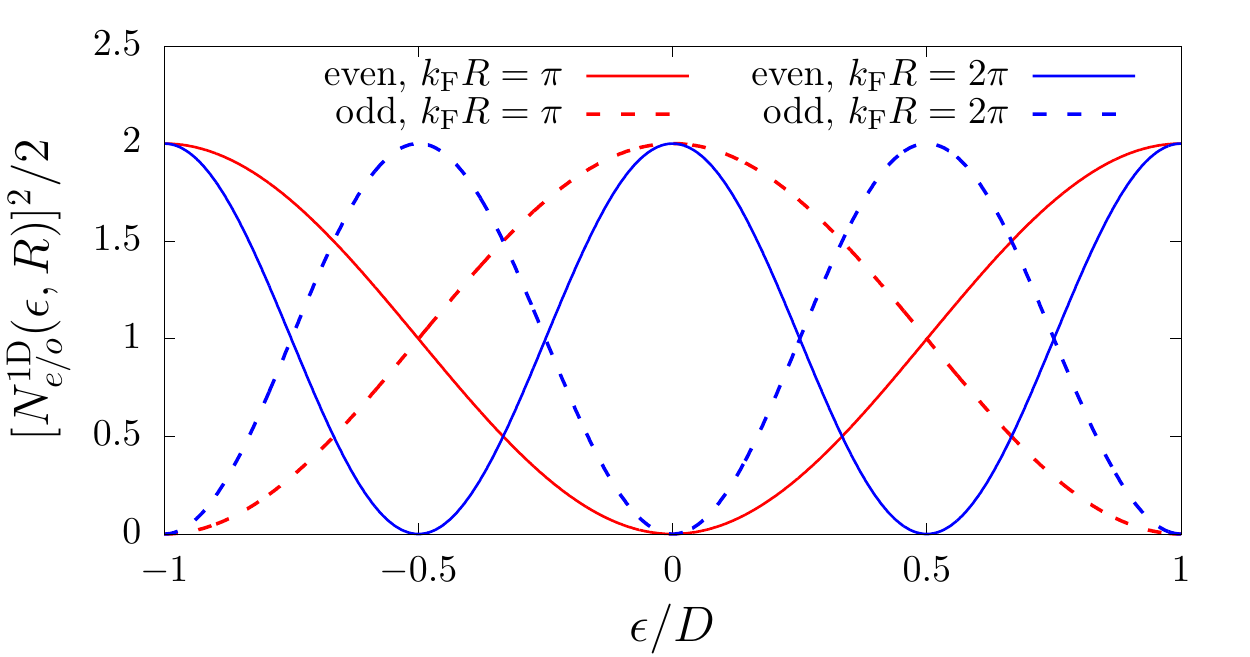}
	\caption{
		Normalization functions of Eq. \eqref{eq:NormalizationFunctions_1D} for a linear dispersion in one dimension for two different distances $k_\mathrm{F}R=\pi$ (red) and $k_\mathrm{F}R=2\pi$ (blue).
		For these distances either the even (solid) or the odd (dashed) normalization function exhibits a pseudo-gap at $\epsilon=0$.
	}
	\label{fig:DOS}
\end{figure}

Up until now we have not specified the dispersion of the conduction-band.
Unless stated otherwise, we will use a 1D linear dispersion $\epsilon(k)=v_\mathrm{F}(|k|-k_\mathrm{F})$ throughout this paper which yields for the normalization functions 
\cite{lit:Borda07,lit:lechtenbergAnders14}
\begin{align}
\left[ N_{e/o}^{\mathrm{1D}}(\epsilon,{R})\right]^2 \rho_c(\epsilon)
=2\rho_c(\epsilon)\left( 1 \pm \cos\left[ k_\mathrm{F} R ( 1 + \frac{\epsilon}{D})\right] \right), \label{eq:NormalizationFunctions_1D}
\end{align}
with the half band width $D=v_\mathrm{F}k_\mathrm{F}$.
$\left[ N_{e/o}^{\mathrm{1D}}(\epsilon,{R})\right]^2$ are plotted for the two different distances $k_\mathrm{F}R/\pi=1$ and $2$ in Fig. \ref{fig:DOS}.
Note that one of the normalization functions exhibits a pseudo-gap at the Fermi energy $\epsilon=0$
for distances $k_\mathrm{F}R/\pi=n$, with $n=0,1,2,\dots$
as a consequence of 
the dispersion $\epsilon(k)=\epsilon(-k)$ \cite{lit:lechtenbergAnders17} employed here. 
This can also be seen from the definitions in Eqs.\ \eqref{eq:NormalizationFunctions_1D}. 
It has been pointed out that the absence of the screening in one of the parity channels
leads to the breakdown of the two-stage Kondo screening process and the 
emergence of a new kind of quantum critical point in the TIKM \cite{lit:lechtenbergAnders17}.
This has also a profound effect on the time dynamics of the TIKM.

In addition to the emergence of the pseudo-gap, both normalization functions are particle-hole symmetric for these special distances and, thus, lead to a completely particle-hole symmetric model.

\subsection{Non-equilibrium dynamics and the TD-NRG}

In order to calculate the real-time dynamics of the TIKM, we employ the TD-NRG, which is an extension of the standard NRG.

The TD-NRG \cite{lit:Anders05,lit:Anders06}
is designed to calculate the full non-equilibrium dynamics of a quantum impurity system after a sudden quench: 
$H(t)=H_0\Theta(-t)+H_f\Theta(t)$.

For this purpose, the initial state of the system is described by the density operator 
\begin{equation}
{\rho}_0 =
     \frac{ e^{-\beta H_0 }}
          {{\rm Tr}
                \left[
                        e^{-\beta H_0 }
                \right] } ,
\label{rho-0}
\end{equation}
until at time $t=0$ the system is suddenly quenched. 
Afterwards, the system is characterized by the Hamiltonian $H_f$ and the time evolution of the density operator is given by
\begin{equation}
{\rho}(t \geq 0) =
    e^{-i t H_f} {\rho}_0
    e^{i t H_f } .
\label{rho-of-t}
\end{equation}
By means of the TD-NRG the time-dependent expectation value $O(t)$ of a general local operator ${O}$ should to be calculated.
In this paper, the local operator is given by the spin-correlation function of both impurities ${O}=\expect{\vec{S}_1\vec{S}_2}$. 

The time evolution of such local operators can be written as \cite{lit:Anders05,lit:Anders06}
\begin{eqnarray}
\langle {O} \rangle (t) &=&
        \sum_{m}^{N}\sum_{r,s}^{\rm trun} \;
        e^{i t (E_{r}^m - E_{s}^m)}
        O_{r,s}^m \rho^{\rm red}_{s,r}(m) ,
\label{eqn:time-evolution-intro} 
\end{eqnarray}
where $E_{r}^m$ and $E_{s}^m$ are the NRG eigenenergies of the Hamiltonian $H_f$ at
iteration $m \le N$, $O_{r,s}^m$ is the matrix
representation of ${O}$ at that iteration,
and $\rho^{\rm red}_{s,r}(m)$ is the reduced density
matrix defined as
\begin{equation}
\rho^{\rm red}_{s,r}(m) = \sum_{e}
          \langle s,e;m|{\rho}_0 |r,e;m \rangle .
\label{eqn:reduced-dm-def}
\end{equation}
in which the environment is traced out.
In Eq. \eqref{eqn:time-evolution-intro} the restricted sums over $r$ and $s$
require that at least one of these states is discarded at iteration $m$.
The temperature $T_N \propto \Lambda^{-N/2}$ 
of the TD-NRG calculation is defined by the length of the NRG Wilson chain $N$
and enters Eq. \eqref{rho-0}.
Here, $\Lambda > 1$ denotes the Wilson discretization parameter.

The TD-NRG comprises two simultaneous NRG runs:
one for the initial Hamiltonian $H_0$ in order to compute
the initial density operator ${\rho}_0$ of the system in Eq. \eqref{rho-0} and
one for $H_f$ to obtain the approximate eigenbasis governing the time evolution in Eq. \eqref{eqn:time-evolution-intro}.

This approach has also been extended to multiple quenches \cite{lit:Costi14}, time evolution of spectral functions \cite{lit:Costi2017} and steady state currents at finite
bias \cite{lit:AndersSSnrg2008,lit:SchmittAnders2009,*lit:SchmittAnders2011,lit:JovchevAnders2013}.
The only error of this method originates from the representation of the bath
continuum by a finite-size Wilson chain \cite{lit:WilsonNRG}.
This error is essentially well understood \cite{lit:Eidelstein2012,lit:Guettge2013} and may lead to artificial oscillations and slight deviations of the long time value from the exact result.
These can be reduced by using an increased number of NRG z-tricks \cite{lit:Yoshida90}.

\section{Equilibrium}
  \label{sec:Equilibrium}

\subsection{Antiferromagnetic coupling $J$}
\label{sec:EquilibriumAntiferromagnetic}

  \begin{figure}[t]
    \centering
    \flushleft{(a)}
    \includegraphics[width=0.5\textwidth]{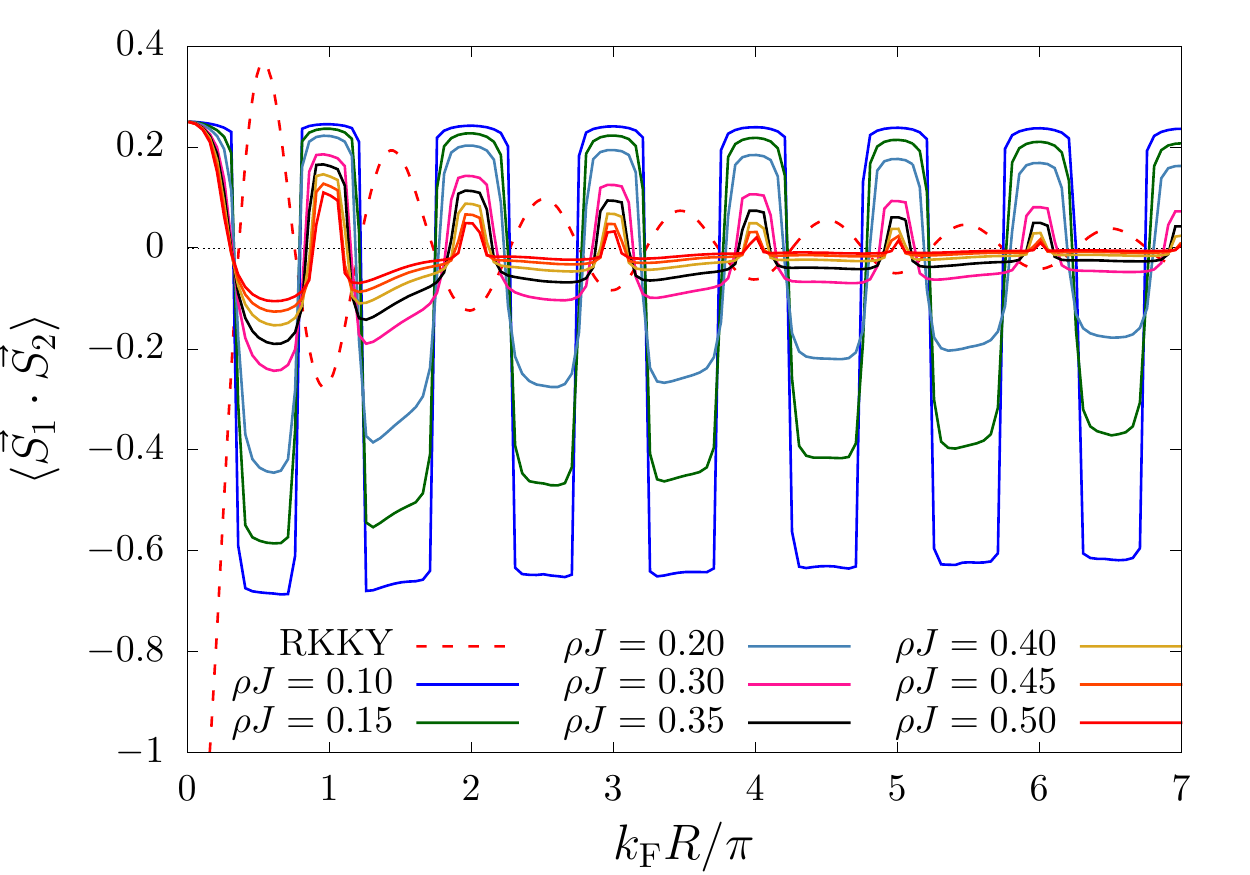}
    \flushleft{(b)}
     \includegraphics[width=0.5\textwidth]{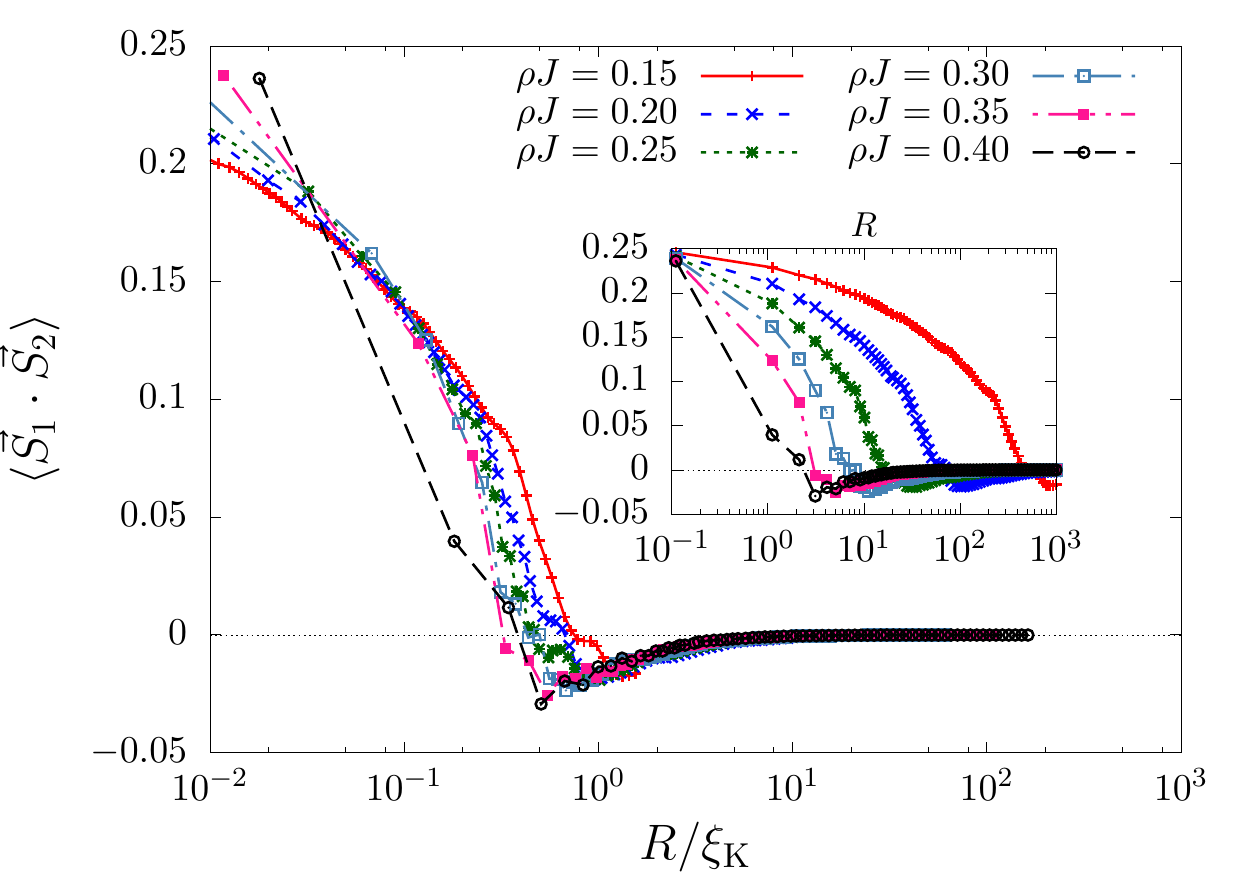}

    \caption{(Color online) 
      (a) The $\langle \vec{S}_1\vec{S}_2\rangle(R)$ correlation function plotted against the distance $k_\mathrm{F}R/\pi$ for different antiferromagnetic couplings $J$.
      The red dashed line depicts the 1D RKKY interaction $\propto 1/R$ in arbitrary units.
      Note that for distances $R \approx \xi_\mathrm{K}$ (large $J$ values) the ferromagnetic correlations around $k_\mathrm{F}R/\pi=n$ begin to vanish.
      For $R \gg \xi_\mathrm{K}$ only at exactly $k_\mathrm{F}R/\pi=n$ ferromagnetic correlations persist.
      (b) 
      Correlation function for the distances $\kf R/\pi=(n+0.11)$ and different couplings $J$ plotted against the rescaled distance $R/\xi_\mathrm{K}$.
      The inset shows the same data vs the distance $R$.
    }
    \label{fig:equi:antiferromagnetic}
  \end{figure}
  
Two characteristic length scales have been identified \cite{lit:Borda07,lit:lechtenbergAnders14}
in the TIKM with an antiferromagnetic $J$ for $T=0$: 
the inverse Fermi momentum $1/\kf$ and the Kondo length scale $\xik=v_\mathrm{F}/\tk$ 
with the Kondo temperature $\tk=\sqrt{\rho J}\mathrm{e}^{-1/{\rho J}}$.
The length scale $1/\kf$ defines the oscillations of the RKKY interaction and its envelope.
As $\xik$ changes exponentially with the Kondo coupling $J$, we use different $J$ to examine 
the different distances $R<\xik$ and $R>\xik$.

The impurity spin-correlation function $\langle \vec{S}_1\vec{S}_2\rangle(R)$ is shown 
in conjunction with the RKKY interaction (dashed line) in 
Fig.\ \ref{fig:equi:antiferromagnetic}(a) for different couplings $J$.
For $R \ll \xi_\mathrm{K}$ (small Kondo couplings $J$) one can clearly observe oscillations between ferromagnetic and antiferromagnetic correlations caused by the RKKY interaction.
For an effective ferromagnetic RKKY interaction, the impurity spins align parallel while for an antiferromagnetic interaction they align antiparallel.

For these small distances $R \ll \xi_\mathrm{K}$ the impurity spins are located inside the respective screening cloud of the other impurity and are not completely screened by the conduction electrons.
Therefore, the Kondo effect has almost no effect on $\langle \vec{S}_1\vec{S}_2\rangle(R)$, which can be seen by a comparison with Fig. \ref{fig:equi:ferromagnetic}(a)
showing $\langle \vec{S}_1\vec{S}_2\rangle(R)$ for ferromagnetic couplings $J$ where the Kondo effect is absent.

Note that $\langle \vec{S}_1\vec{S}_2\rangle(R)$
does not decay  but instead exhibits step like oscillations with a constant amplitude 
since it reflects the ground-state properties of
two free impurity spins which are coupled via a Heisenberg interaction.
Even for an infinitesimal small effective Heisenberg interaction between the impurity spins, the spins align completely parallel or antiparallel at $T=0$.

This behavior is modified for larger couplings $J$, where $R\ll \xi_K$ is not valid anymore due to an increasing Kondo temperature.
Upon lowering the temperatures, the spins begin to align parallel or antiparallel until the Kondo temperature $\tk$ is reached,
at which the impurities are screened by the conduction electrons and, hence, the correlation function does not change anymore.
The exact value of the correlation function depends on the ratio between the RKKY interaction and the Kondo temperature $K_\mathrm{RKKY}/\tk$ \cite{lit:Jones88}.

Furthermore, one should note that when the distance approaches the Kondo length scale,
$R \sim \xik$, large $J$ and $R$ in Fig.\ \ref{fig:equi:antiferromagnetic}(a),
the Kondo effect leads to a drastic departure from the conventional RKKY interaction \cite{lit:Feiguin2015}.
While for small $J$ and $R$ the position of the 
sign change
of the RKKY interaction agrees with the position of the sign change
of the correlation function,
the latter is shifted towards the integer distances $\kf R/\pi=n$ with increasing coupling $J$ and distance $R$.
The interval in the vicinity of 
the distances $\kf R/\pi=n$ where we observe ferromagnetic correlation, therefore, shrinks and
antiferromagnetic correlations between the impurity spins emerge instead\cite{lit:Fye1989}.

Since $\xik$ exponentially depends on the Kondo coupling $J$, the precise distance $R$ at 
which the ferromagnetic correlations disappear is also $J$ dependent.
Therefore,
one almost only observes antiferromagnetic correlations between the impurities
for $R\gg \xik$.

In order to review the influence of the Kondo effect on the ferromagnetic correlations,
we calculated $\langle \vec{S}_1\vec{S}_2\rangle(R)$ at the distances $\kf R/\pi=(n+0.11)$ 
where we expect a finite FM RKKY interaction. The results are shown in Fig. \ref{fig:equi:antiferromagnetic}(b)
plotted as a function of the rescaled distance $R/\xik$ and as a function of $R$ in the inset.
The crossover from FM to AFM is governed by the Kondo effect
and occurs once the distance exceeds $R\sim 0.53 \xi_\mathrm{K}$.

Based on the observed universality,  we can understand this surprising 
sign change of the spin-correlation function 
within the strong coupling limit. For $J\to \infty$, a Kondo singlet is formed locally at each impurity site,
and the local conduction-band electron is antiparallel
to the local spin. 
In this case, the system consists of two Kondo singlets which are decoupled from the remaining Fermi sea with two missing electrons.
In the generic case, however, 
the two bound states in the even-odd basis 
are subject to
the potential scattering terms emerging from the particle-hole asymmetry 
(see Sec.\ \ref{sec:ModelMethod:Mapping}).
These scattering terms are different for the even and odd conduction-band channel 
\cite{lit:Varma2007,lit:Affleck95} and, hence, 
generate a hopping term between the bound states in the real-space basis.

This hopping term evokes an antiferromagnetic interaction so that 
the two bound conduction electron spins arrange in opposite orientation
inducing an AF correlation between the impurity spins
as observed in Fig.\ \ref{fig:equi:antiferromagnetic}(b) for $R/\xik > 1$.

A word is in order to justify the choice $\kf R/\pi=(n+0.11)$ as generic distance.
$\kf R/\pi=n$ leads to a different physics \cite{lit:lechtenbergAnders17} for a linear dispersion in one dimension
considered here for two reasons:
At first, one of the two parity
conduction bands develops a pseudo-gap DOS at low temperatures, as depicted in Fig.\ \ref{fig:DOS},
and does not participate in the screening any more.
Second, at $\kf R/\pi=n$ the system is perfectly particle-hole symmetric and the above-mentioned additional hopping term between the bound conduction electrons does not appear.
Consequently, the system is equivalent to the physics
at $R=0$ \cite{lit:lechtenberg_2016_dimer} for these distances and ferromagnetic correlations remain for all integers $n$.

\begin{figure}[t]
    \centering
    \includegraphics[width=0.5\textwidth]{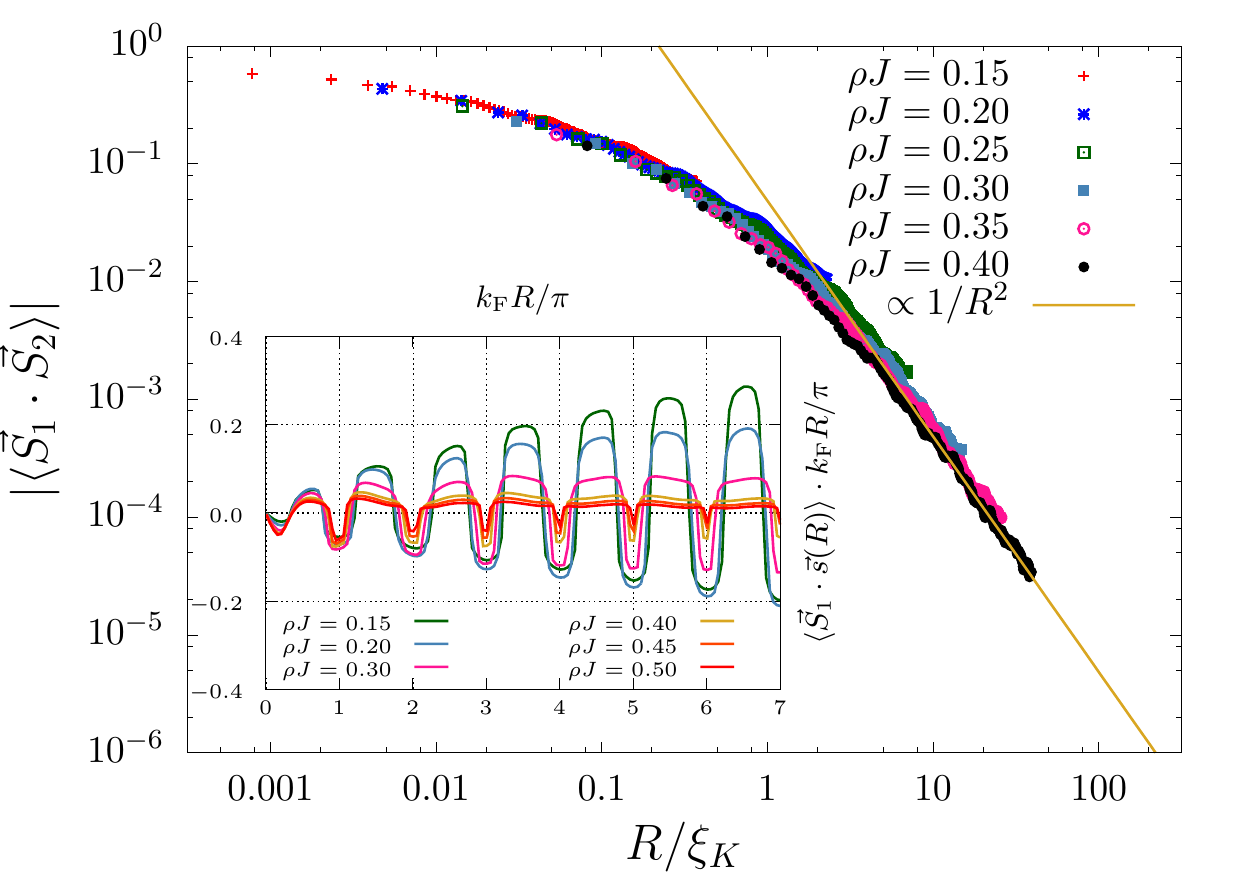}
    \caption{(Color online) 
      The envelope $|\langle \vec{S}_1\vec{S}_2\rangle(R)|$ of the impurity spin-correlation function on a double logarithmic scale plotted against the rescaled distance $R/\xi_\mathrm{K}$.
      The rescaling leads to a universal behavior.
      For large distances $R \gg \xi_\mathrm{K}$ a $1/R^2$ decrease is observed.
      The inset shows the correlation function $\langle \vec{S}\vec{s}(R)\rangle$ between an impurity spin $\vec{S}$ 
      and the conduction-band spin density $\vec{s}(R)$ at distance $R$ from the impurity.
    }
    \label{fig:equi:antiferromagnetic_envelope}
\end{figure}

A similar behavior has also been observed in the single impurity Kondo model (SIKM) for the correlation function $\langle \vec{S}\vec{s}(R)\rangle$ which
measures the correlations between the impurity spin and the spin density of the conduction band in distance $R$ to the impurity \cite{lit:lechtenbergAnders14}.
The ferromagnetic correlations located at $\kf R/\pi=(n+0.5)$ vanish for distances $R > \xik$ 
and instead also antiferromagnetic correlations appear
in accordance with theoretical predictions \cite{lit:Barzykin1998,lit:Ishii1978}.

The inset of Fig. \ref{fig:equi:antiferromagnetic_envelope} shows the same correlation function $\langle \vec{S}\vec{s}(R)\rangle$ for the TIKM
measuring the correlation between an impurity spin and the conduction-band spin density at the position of the second impurity located a distance $R$ from the first impurity.
To counteract the decay, the correlation function has been rescaled with the distance $R$ for a better prospect.
In comparison to the correlation function for the SIKM, 
the second impurity leads to a $\pi/2$ phase shift such that now the antiferromagnetic correlations around $\kf R/\pi=n$ instead of the ferromagnetic ones around $\kf R/\pi=(n+0.5)$ vanish.
Consequently, the ferromagnetic correlations between the impurity spins $\langle \vec{S}_1\vec{S}_2\rangle(R)$ at the distances $\kf R/\pi=n$ also have to vanish 
since the RKKY interaction between the impurity spins is mediated by the conduction band.

Figure \ref{fig:equi:antiferromagnetic_envelope} depicts the envelope of $\langle \vec{S}_1\vec{S}_2\rangle(R)$ measured at the distances $\kf R/\pi=(n+0.5)$.
The universal behavior of the envelope function is revealed by plotting the data as a function of the dimensionless distance $R/\xik$.
This shows that the amplitude of the correlation function is completely governed by the distance dependent RKKY interaction and the Kondo effect.
For large distances $R \gg \xik$ a $\propto 1/R^2$ behavior, indicated by the solid line, is observed.
At these large distances the impurities are located outside of the Kondo screening cloud of the respective other almost completely screened impurity,
therefore, the $\propto 1/R$ decay of the RKKY interaction in one dimension is enhanced to a $\propto 1/R^2$ decay.
The same $\propto 1/R^2$ behavior for $R \gg \xik$ has also been found for the correlation 
between an impurity spin and the conduction-band spin density $\langle \vec{S}\vec{s}(R)\rangle$ in the SIKM \cite{lit:lechtenbergAnders14}.

\subsection{Ferromagnetic coupling $J$ and finite temperatures}
  \label{sec:EquilibriumFerromagnetic}
  
\begin{figure}[t]
\centering
 \flushleft{(a)}
 \includegraphics[width=0.5\textwidth]{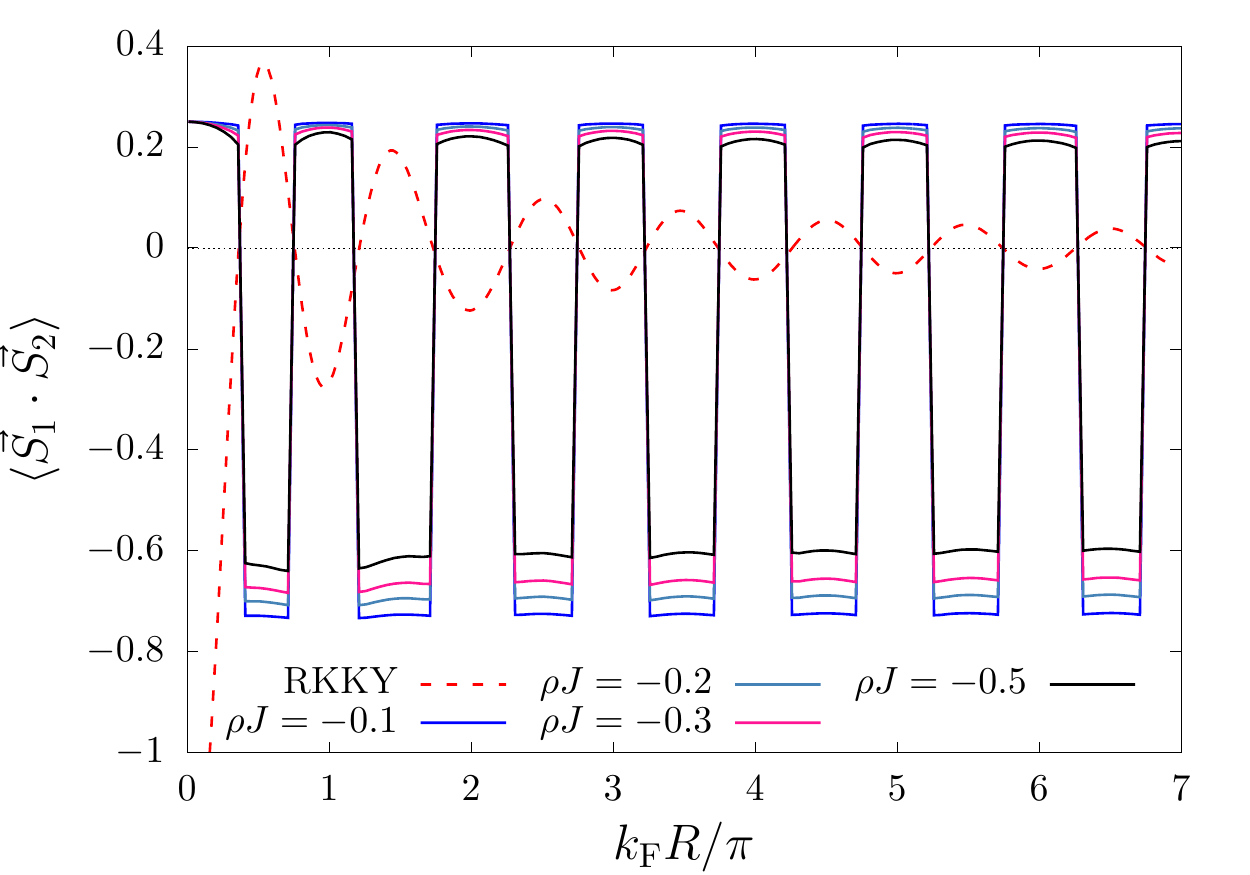}
 \flushleft{(b)}
    \includegraphics[width=0.5\textwidth]{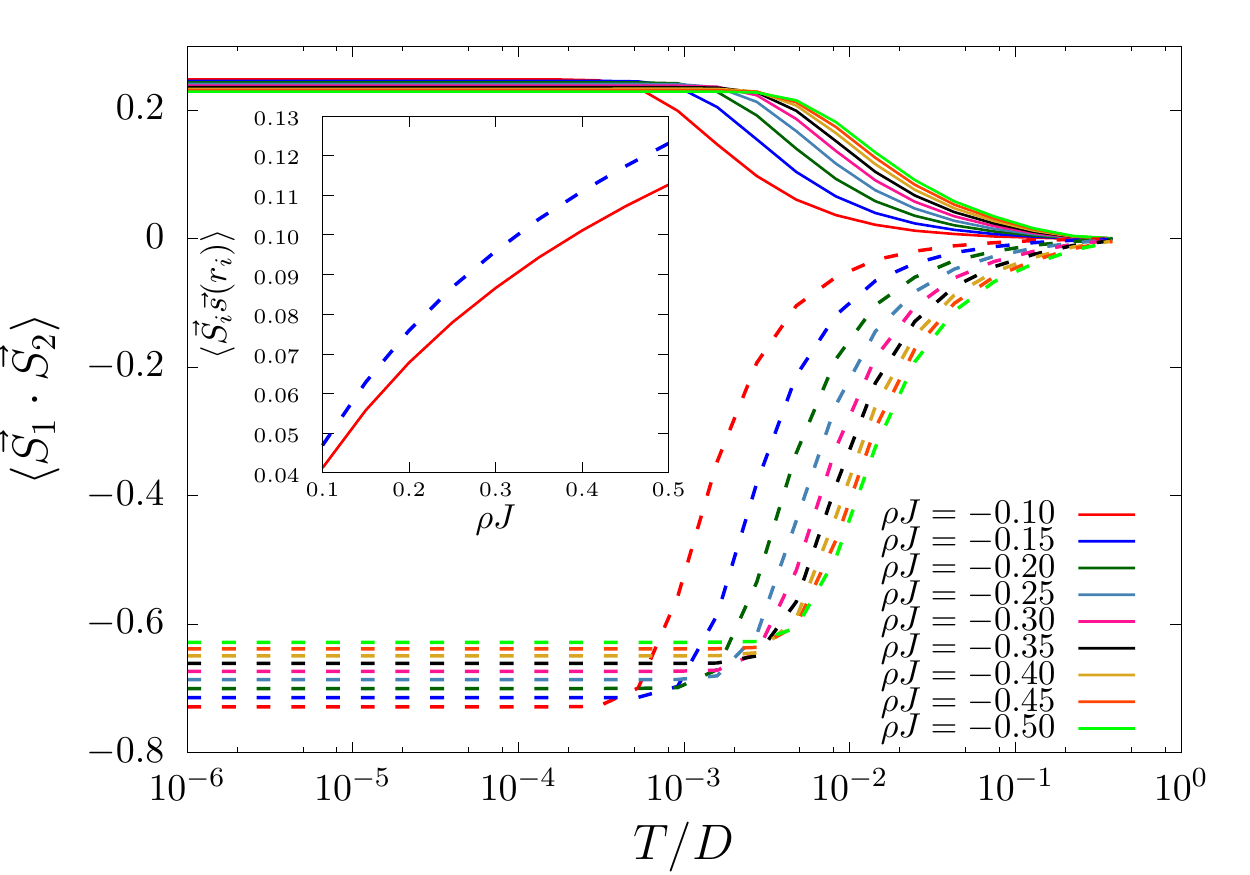}

\caption{(Color online) 
(a) The $\langle \vec{S}_1\vec{S}_2\rangle(R)$ correlation function vs the distance $k_\mathrm{F}R/\pi$ for different ferromagnetic couplings $J$.
The red dashed lines depicts the 1D RKKY interaction $\propto 1/R$ in arbitrary units.
(b) Temperature-dependent correlation function for different couplings and the two different distances $\kf R/\pi=1.0$ (solid lines), where the RKKY interaction is ferromagnetic,
and $\kf R/\pi=0.5$ (dashed lines), where the RKKY interaction is antiferromagnetic.
The inset shows the fixed-point value of the correlation between an impurity spin and the spin density of the conduction electrons
at the position of this impurity $\langle\vec{S}_i\vec{s}(r_i)\rangle$ for different couplings $J$ and the two distances $\kf R/\pi=1.0$ (red solid line)
and $\kf R/\pi=0.5$ (blue dashed line).
}
\label{fig:equi:ferromagnetic}
\end{figure}
So far, we have only investigated the TIKM for an antiferromagnetic coupling $J$ where the Kondo effect is present.
We now extend our discussion also to ferromagnetic $J$.

The correlation function $\langle \vec{S}_1\vec{S}_2\rangle(R)$ for
ferromagnetic couplings as well as the RKKY interaction (dashed line) is depicted in Fig. \ref{fig:equi:ferromagnetic}(a). 
Since the Kondo effect is absent, there is no
screening of the local moments with increasing $J$
in contrast to AFM $J$ shown in  Fig.\ \ref{fig:equi:antiferromagnetic}(a).
The correlation function preserves its step like oscillations even for very large ferromagnetic couplings.

Similar to the case for antiferromagnetic $J$, we observe 
that the modulus of $\langle \vec{S}_1\vec{S}_2\rangle(R)$ is reduced with increasing $|J|$.
However, the decrease is much weaker than for antiferromagnetic $J$. 
This reduction cannot be caused by a screening of the impurity and, therefore, must have a different origin.

\begin{figure}[t]
\centering
 \includegraphics[width=0.5\textwidth]{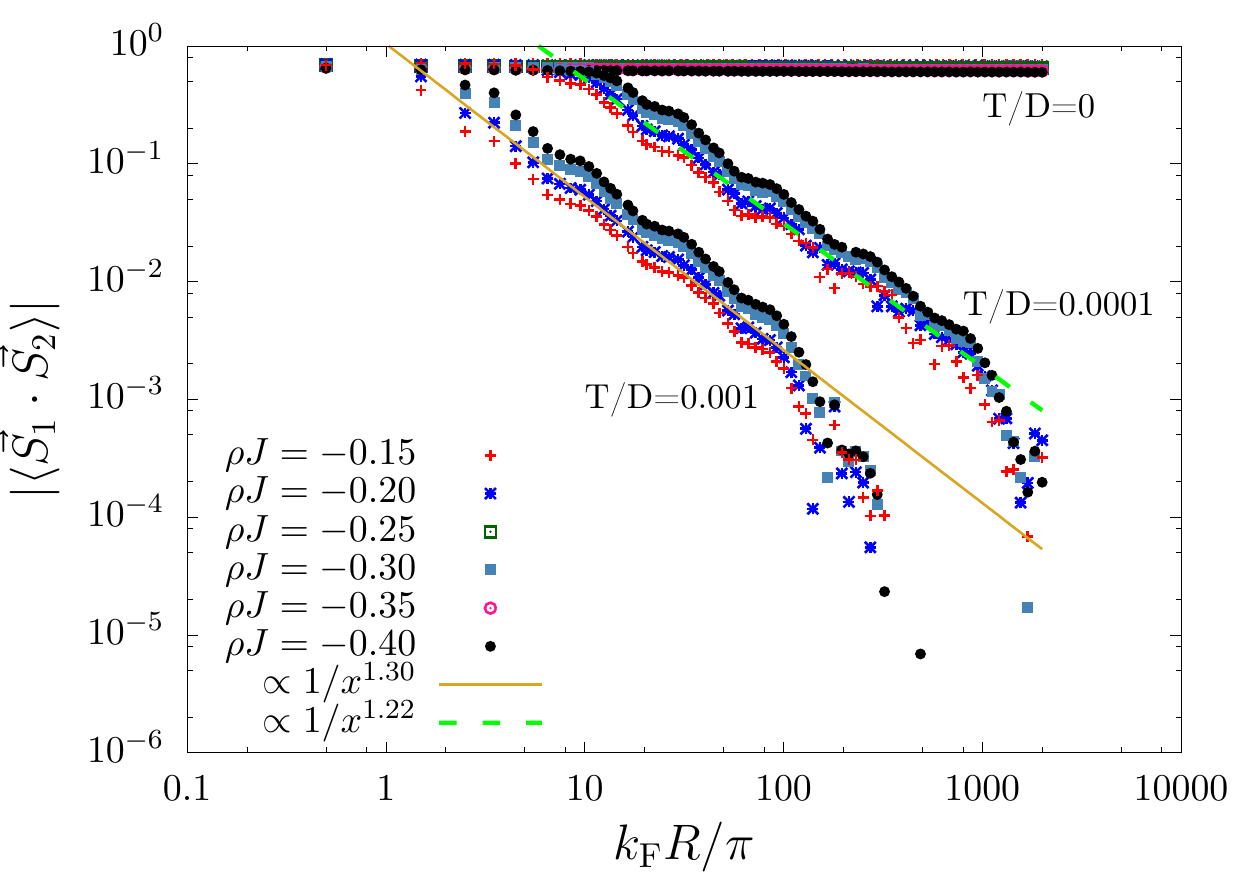}
\caption{(Color online) 
The envelope of $\langle \vec{S}_1\vec{S}_2\rangle(R)$ for ferromagnetic couplings depicted on a double logarithmic scale.
For $T=0$, the amplitude of the correlation function remains constant for all distances.
For the finite temperatures $T/D=0.001$ and $0.0001$ a power-law decay is observed when the RKKY interaction is smaller than the temperature T 
which turns over into an exponential decay once the length scale $\xi_\mathrm{T}=v_\mathrm{F}/T$ is reached.
}
\label{fig:equi:ferromagnetic_Envelope}
\end{figure}

Figure \ref{fig:equi:ferromagnetic}(b) depicts the temperature dependent
correlation function for different ferromagnetic couplings $J$ and for the distance $\kf R/\pi=1.0$ (solid lines), where the RKKY interaction is ferromagnetic,
as well as for $\kf R/\pi=0.5$ (dashed lines), where the RKKY interaction is antiferromagnetic.
For both regimes we observe (i) a reduction of the modulus of the spin-correlation function with
increasing Kondo coupling $J$
and (ii) an simultaneous increase of the crossover temperature from 
two uncorrelated spins at high temperatures to spin correlation in the fixed-point.

For a ferromagnetic coupling $J<0$, the effective coupling in the NRG renormalization flow is renormalized to zero $J_\mathrm{eff}\to 0$ for $T\to0$ \cite{lit:PWAnderson1970_II}.
As soon as the effective coupling is zero, a fixed point is reached and, consequently,
the correlation function reaches its fixed point value.
Note, however, that the operator content of the renormalized operators is important: The larger the
Kondo coupling, the smaller the fraction of the original spin that contributes to the effective spin degree of freedom
that decouples from the conduction band.
This is demonstrated in the inset of Fig. \ref{fig:equi:ferromagnetic}(b), which shows the fixed-point value of the correlation between an impurity spin and the spin density of the conduction electrons
at the position of this impurity $\langle\vec{S}_i\vec{s}(r_i)\rangle$ for different couplings $J$.
The correlations remain finite in the fixed point even if the effective coupling is renormalized to zero  since only a part of the impurity spins decouples.
The fraction of the impurity spins which remains coupled to the conduction band is the larger the larger $J$ is.
\\
Therefore, for a finite coupling to the conduction band the effective decoupled spins are reduced in the renormalization flow 
until the fixed point is reached where $J_\mathrm{eff}=0$ 
which
is the origin of the reduction of
$|\langle \vec{S}_1\vec{S}_2\rangle(R)|$ for FM and AFM RKKY couplings. 

Note, however, that a clearly noticeable reduction of the amplitude occurs only for very large ferromagnetic couplings $J$.

An increasing crossover scale to the fixed point with increasing $J$ is not observed in the SIKM and can, therefore, be ascribed to a growing RKKY interaction $\propto J^2$
since it is the only additional effect in a TIKM with ferromagnetic couplings.
We have also checked that an increasing direct Heisenberg interaction between the impurity spins, as given in Eq. \eqref{eq:H_Imp}, with vanishing RKKY interaction ($R\to\infty$)
has the same effect as a finite indirect RKKY interaction and also leads to an increasing crossover scale.
It is already known that the RKKY interaction has a profound effect on the renormalization flow of the TIKM for antiferromagnetic Kondo couplings \cite{lit:Kroha17}.

Figure \ref{fig:equi:ferromagnetic_Envelope} shows the envelope of the correlation function for different ferromagnetic couplings and different temperatures.
As can be seen, for zero temperature $T/D=0$, the amplitude is almost constant even for $R \to \infty$.
This changes for the finite temperatures $T/D=0.001$ and $0.0001$ where a power-law decay is observed as soon as the energy scale of the RKKY interaction is smaller than the temperature.
However, the finite temperature also introduces a new length scale $\xi_\mathrm{T}=v_\mathrm{F}/T$ beyond which the correlation function decays exponentially.
The same finite temperature behavior has also been found in the SIKM for the correlation between the impurity spin and the spin density of the conduction band 
at a distance $R$ from the impurity \cite{lit:Borda07}.

\section{Real-time dynamics of the TIKM}
\label{sec:Nonequilibrium}

\subsection{Spin-correlation function after a quench}

\begin{figure}[t]
	\centering
  \flushleft{(a)}
	\includegraphics[width=0.5\textwidth]{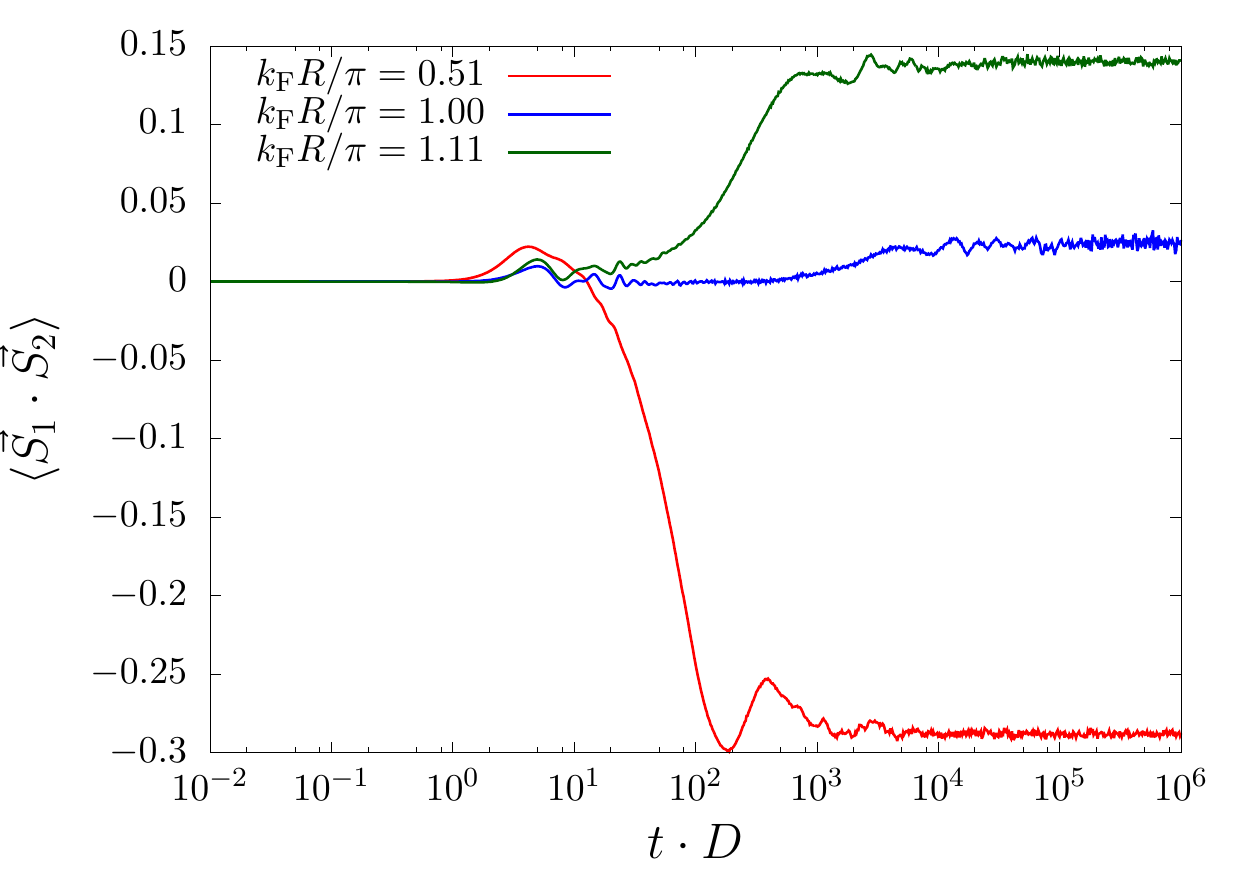}
  \flushleft{(b)}
	\includegraphics[width=0.5\textwidth]{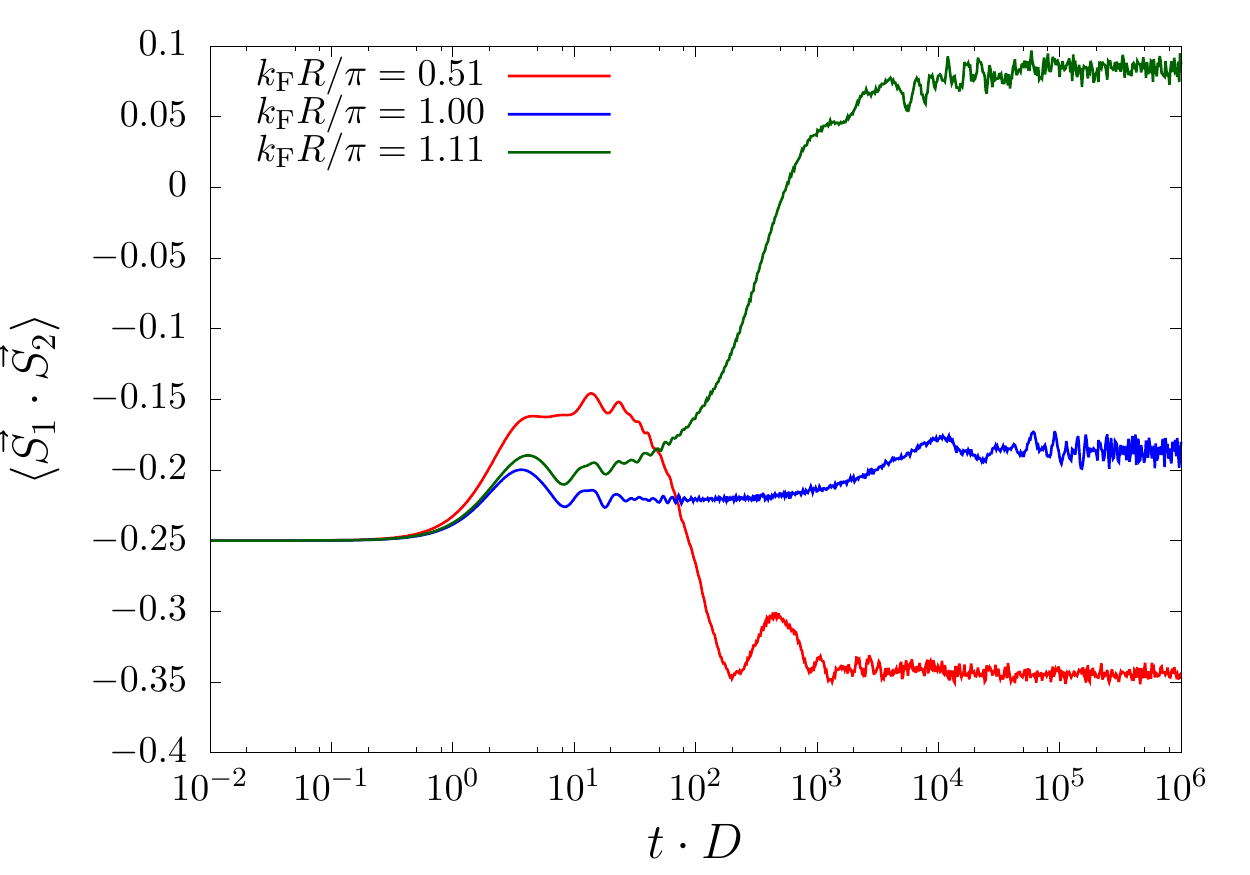}
\caption{(Color online) 
	(a) The long-time behavior of $\langle \vec{S}_1\vec{S}_2\rangle(R,t)$ after a quench in the coupling from $\rho J=0$ to
	$0.2$ for the three different distances $\kf R/\pi=0.51$, $1.00$ and $1.11$.
	(b) Time dynamics of $\langle \vec{S}_1\vec{S}_2\rangle(R,t)$ after a quench in magnetic fields applied to the impurities from $H_1=-H_2=10D$ to $H_1=H_2=0$
	for the same distances as in (a).
	NRG parameters: $\lambda=3$, $\mathrm{Ns}=2000$ and $N_z=32$. 
}
\label{fig:longtime_different_distances}
\end{figure}

The discussion of the  equilibrium properties in the previous section
sets the stage for the investigation of the real-time dynamics 
of the time-dependent spin-correlation function $\langle \vec{S}_1\vec{S}_2\rangle(R,t)$
after a quench of the system.
We focus on antiferromagnetic Kondo couplings $J>0$
and set the coupling of the impurities to the conduction band initially to zero $J=0$
such that the impurities are completely decoupled from the band.
At time $t=0$ the coupling is switched on to a finite antiferromagnetic value $J > 0$ and the time-dependent behavior of $\langle \vec{S}_1\vec{S}_2\rangle(R,t)$
is calculated using the TD-NRG by evaluating Eq.\ \eqref{eqn:time-evolution-intro}.

Figure \ref{fig:longtime_different_distances}(a) shows
$\langle \vec{S}_1\vec{S}_2\rangle(R,t)$ for times up to $tD=10^6$ after such a quench
for three different distances. 
The RKKY interaction is antiferromagnetic at the distance $\kf R/\pi=0.51$ and 
ferromagnetic for the distances $\kf R/\pi=1.00$ and $1.11$.
As can be seen, the correlation function behaves very differently for the three 
different distances, even for the two distances at which the RKKY interaction is ferromagnetic
and a similar behavior is expected.

A ferromagnetic correlation emerges for small times for
all distances the origin of which is caused by a ferromagnetic wave propagating through the system as we will show later. 
For the distance $\kf R/\pi=0.51$, the correlation function
becomes antiferromagnetic only  at longer times and approaches its equilibrium value. 
Note the log timescale in Fig.\  \ref{fig:longtime_different_distances}(a).
The equilibrium value of about $\langle \vec{S}_1\vec{S}_2\rangle(\kf R/\pi=0.51)\approx -0.42$ 
is, however, not completely reached. 
For strong antiferromagnetic interactions the two impurity spins form a singlet and thus decouple from the conduction band \cite{lit:Jones88,lit:Jones1989}.
Without any additional relaxation mechanism, this decoupling prevents the correlation function from reaching its equilibrium value.

The RKKY interaction has a ferromagnetic maximum for $\kf R/\pi=1.00$.
Strikingly, the correlation function changes only for short times and remains almost constant after the first ferromagnetic maximum. 
This surprising behavior is related to the property of the dispersion, i.\ e.\
$\epsilon(k)=\epsilon(|k|)$  \cite{lit:lechtenbergAnders17}.
At special distances $\kf R/\pi=n$, we observe that the impurity correlation function 
$\langle \vec{S}_1\vec{S}_2\rangle$ 
becomes a conserved quantity
resulting in a fixed value for $\langle \vec{S}_1\vec{S}_2\rangle(R,t)$ for long times.

In order to understand this effect, one has to examine the energy dependent normalization functions between the impurities and the conduction band.
For a 1D 
symmetric dispersion, either the even or the odd normalization function in Eq. \eqref{eq:NormalizationFunctions_1D} exhibits a pseudo-gap at the Fermi energy $\epsilon=0$
for the distances $\kf R/\pi=n$, with $n=0,1,2,\dots$ (see also Fig. \ref{fig:DOS}).
Due to the pseudo-gap either $N_e(0,R)=0$ or $N_o(0,R)=0$ always vanishes at the Fermi energy for these special distances.
This also leads to the fact that the last term of the Hamiltonian in Eq. \eqref{eq:TIKM:H_I} proportional to $\propto (\vec{S}_1 -\vec{S}_2) N_e(\epsilon,R)N_o(\epsilon',R)$
always vanishes on low energy scales for the distances $\kf R/\pi=n$.
This term is, however, responsible for the correlation function to smoothly evolve from a spin triplet to a singlet value or vice versa
since it mixes electrons from the even and odd conduction band via impurity scattering processes. 
In a parity-symmetric TIKM the global parity remains conserved; however, the local impurity parity and the parity in the conduction bands may change.
Once this term vanishes, the band mixing is suppressed and, therefore, the local impurity parity becomes a conserved quantity at low-energy scales.
Consequently, the correlation function $\langle \vec{S}_1\vec{S}_2\rangle(\kf R/\pi=n,t)$ is fixed for long times due to parity symmetry.

Note that this effect is not necessarily restricted to 1D dispersions.
Generally, a dispersion is needed where at certain distances either the even or the odd normalization function in Eq. \eqref{eq:NormFactor} vanishes or, at least, almost vanishes for small temperatures
inducing a local parity conservation.

At the distance $\kf R/\pi=1.11$ the RKKY interaction is also ferromagnetic,
but the effective density of states does not exhibit a pseudo-gap. 
$\langle \vec{S}_1\vec{S}_2\rangle(R,t)$ approaches its ferromagnetic equilibrium value, as expected.
However, the equilibrium value of $\langle \vec{S}_1\vec{S}_2\rangle(R)\approx 0.2$ is  not completely reached.

Although qualitatively the results remain unchanged, 
the long-time limit of $\langle \vec{S}_1\vec{S}_2\rangle(R,t)$ slightly depends on the discretization parameter $\Lambda$ of the NRG for times $t D> 1000$.

In order to demonstrate that the characteristic difference in the real-time dynamics
of the correlation function is not only restricted to quenches in the coupling $J$,
Fig. \ref{fig:longtime_different_distances}(b) shows the behavior of $\langle \vec{S}_1\vec{S}_2\rangle(R,t)$ 
after a quench in magnetic fields applied to the impurity spins from $H_1=-H_2=10D$ to $H_1=H_2=0$.
Since the impurity spins are initially antiparallel aligned, the correlation function starts from $\langle \vec{S}_1\vec{S}_2\rangle(R,0)=-0.25$ at $t=0$.
As can be seen, the behavior is very similar to Fig. \ref{fig:longtime_different_distances}(a) such that for the distances $\kf R/\pi=0.51$ and $1.11$ the correlation function approaches 
its equilibrium value while for $\kf R/\pi=1.00$ it remains close to its initial value.
Also note that the initial condition with antiparallel aligned impurity spins is not parity symmetric; however, the Hamiltonian driving the time dynamics is parity symmetric and, therefore, 
leads to a local impurity parity conservation for the distance $\kf R/\pi=1.00$ for long times.

\subsection{Short-time behavior}
\label{sec:Nonequilibrium_Short}
\begin{figure}[t]
\centering
 \flushleft{(a)}
 \includegraphics[width=0.5\textwidth]{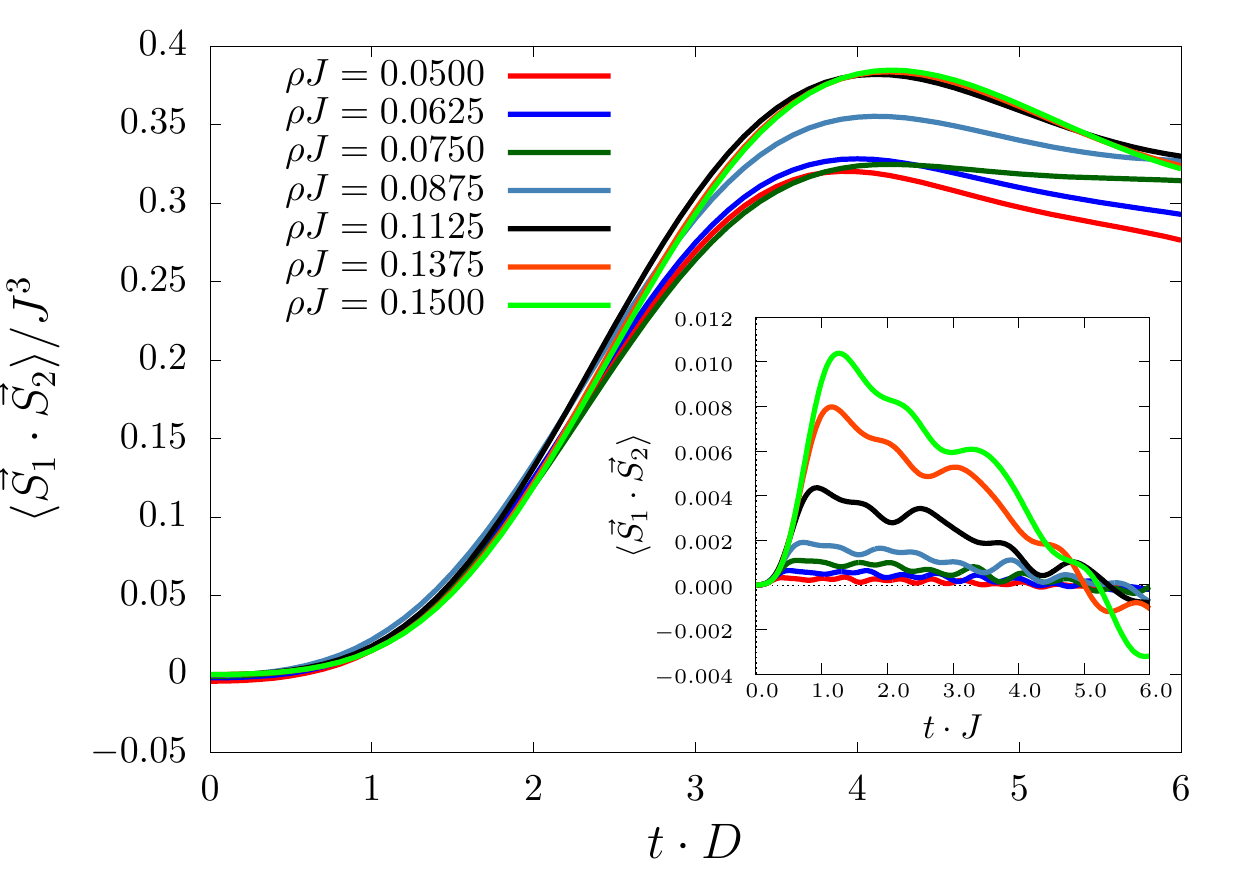}
 \flushleft{(b)}
 \includegraphics[width=0.5\textwidth]{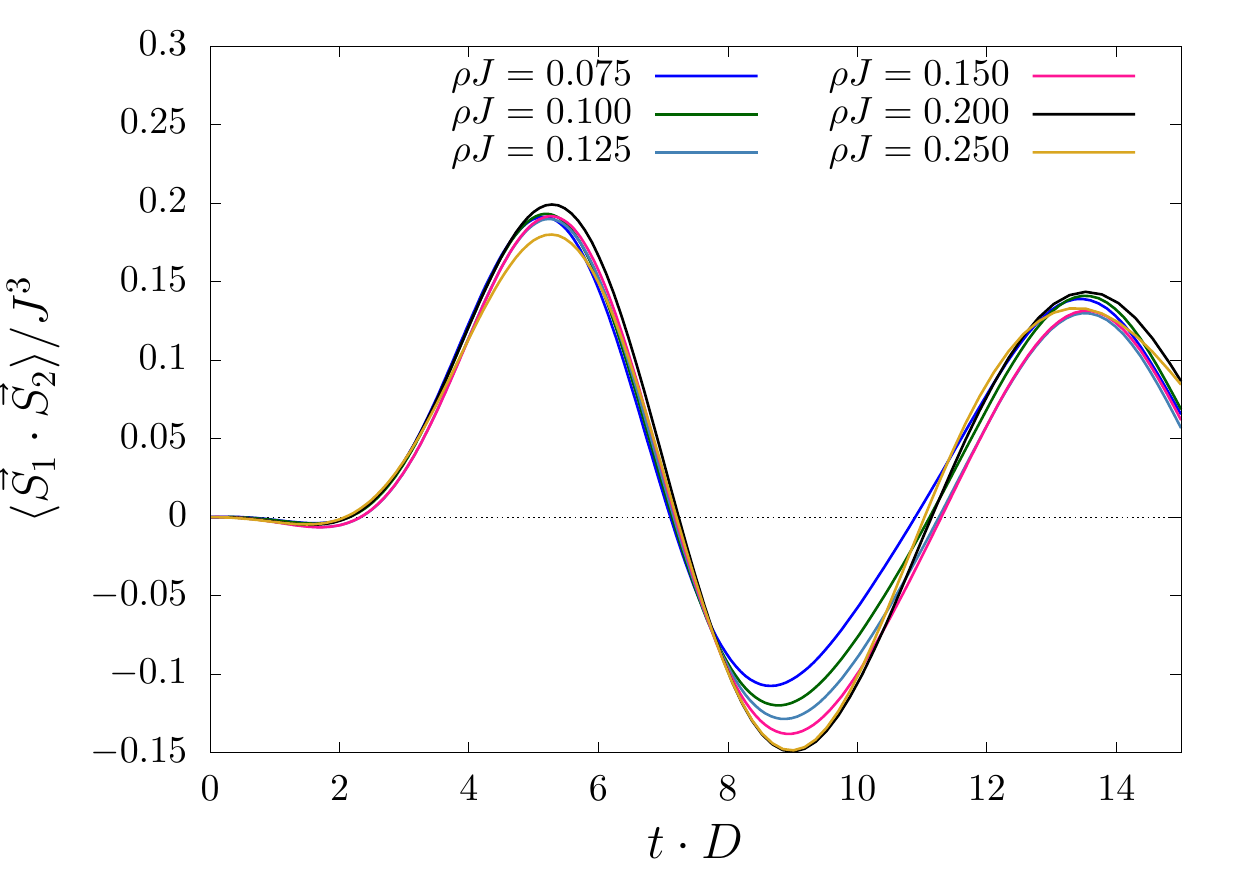}

\caption{(Color online) 
(a) The short time behavior of the spin-correlation function of the TIKM rescaled with $1/J^3$ for different couplings $J$ and the fixed distance $k_\mathrm{F}R=0.51\pi$.
The inset depicts $\langle \vec{S}_1\vec{S}_2\rangle(R,t)$ against the rescaled time $t J$.
Note that, due to the rescaling with $J$, the zero crossings from positive to negative correlations at $t J \approx 5$ approximately coincide for all $J$.
(b) Short-time behavior for the distance $k_\mathrm{F}R=1.00\pi$ and different couplings $J$ plotted against $t D$.
For this distance the zero crossings from positive to negative correlations at $t D\approx 7$ coincide without any rescaling of the time.
NRG parameters: $\lambda=3$, $\mathrm{Ns}=2000$, $N_z=16$.
}
\label{fig:shorttime}
\end{figure}

After presenting the real-time dynamics for all timescales in the previous
section, we now discuss the short-time behavior in more detail.

For zero initial correlation function $\langle \vec{S}_1\vec{S}_2\rangle(R,0)=0$,
the first- and second-order contributions in a perturbation expansion in $J$ vanish 
so that the first non-vanishing order is $\propto J^3$
(for details see Appendix \ref{app:perturbation_theory}).
The impurity correlation function $\langle \vec{S}_1\vec{S}_2\rangle(R,t)$
calculated  with the TD-NRG 
is depicted in Fig. \ref{fig:shorttime}(a) for the distance $k_\mathrm{F}R=0.51\pi$ 
and different antiferromagnetic couplings $J$.
By rescaling the results with $1/J^3$  we demonstrate a perfect agreement with the
scaling prediction of the perturbation theory which becomes exact in the limit $t\to 0$.

Around this distance a ferromagnetic correlation 
develops where the peak position is only dependent on the reciprocal band width,
and, therefore, related to the Fermi velocity. We will show below, that this peak will be linearly dependent on the distance
$R$ between the impurities, and is related to the information spread between the two impurities.

The inset of Fig. \ref{fig:shorttime}(a) shows the correlation function for the same distance and couplings plotted against the rescaled time $t J$.
For the increase of the correlation function at times $t J < 1$ again a universal short-time behavior is found.
We can, therefore, conclude that the initial build up of the ferromagnetic wave is proportional to $\propto (t J)^3$.

For the distances $k_\mathrm{F}R/\pi=n + 0.5$ the equilibrium correlation function is antiferromagnetic since 
the RKKY interaction reaches its largest antiferromagnetic amplitude during each oscillation cycle.
However, $\langle \vec{S}_1\vec{S}_2\rangle(R,t)$ remains ferromagnetic for a relatively long time
before it later approaches its antiferromagnetic long-time value.
The inset of Fig. \ref{fig:shorttime}(a) reveals that the timescale of this ferromagnetic range is given by $1/J$ since for the rescaled time $t J$
the zero crossing from ferromagnetic to antiferromagnetic correlations is approximately $t J \approx 5$ for all couplings $J$.

Figure \ref{fig:shorttime}(b) depicts the rescaled correlation function $\langle \vec{S}_1\vec{S}_2\rangle(R,t)/J^3$ for the distance $\kf R=1.00\pi$ and different couplings.
As before, a universal build up of the ferromagnetic wave can be observed.

For this distance, however, the sign change from ferromagnetic to antiferromagnetic correlations is governed by the inverse
band width $D$ that is proportional to the Fermi velocity. 
Because of the decoupling of one effective band, the local parity is dynamically conserved.
The energy scale, and consequently also the timescale, of the decoupling are defined by the distance between the impurities and the bandwidth of the conduction band $D$.
Therefore, due to the pseudo-gap formation at the distances $k_\mathrm{F}R/\pi=n$,
the relevant timescale is given by $D$.

\subsection{Long-time behavior}
\label{sec:Nonequilibrium_Long}

\begin{figure}[t]
	\centering
  \flushleft{(a)}
	\includegraphics[width=0.5\textwidth]{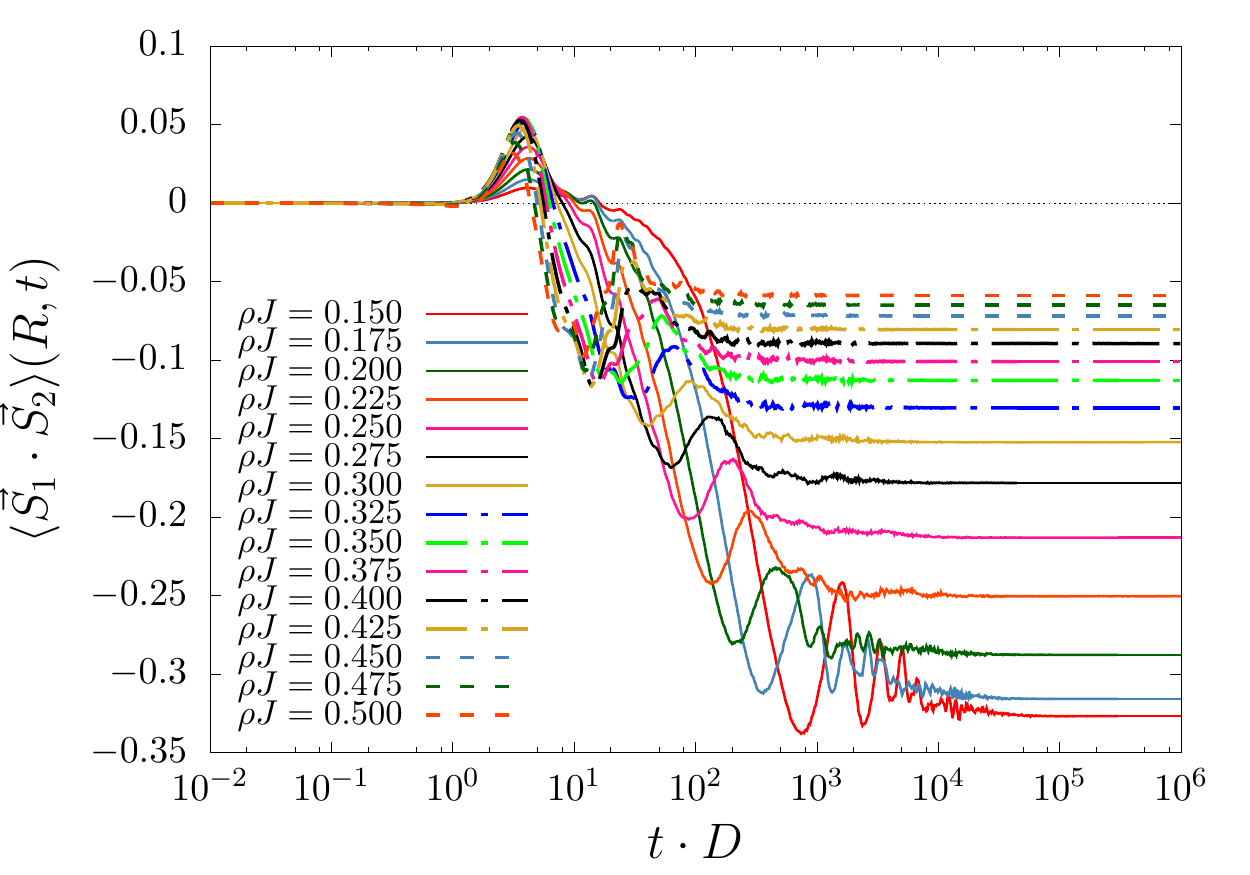}
  \flushleft{(b)}
  \includegraphics[width=0.5\textwidth]{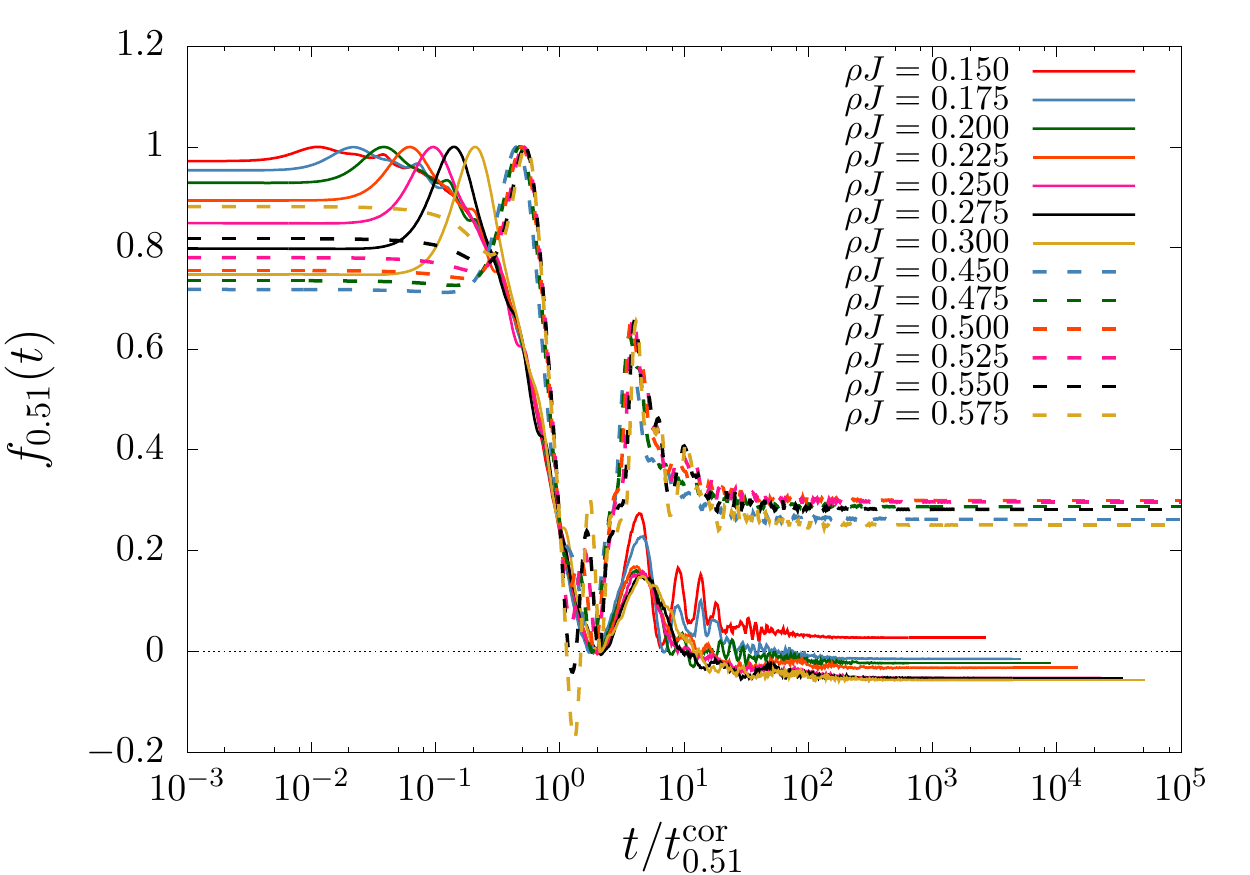}
\caption{(Color online) 
	(a) $\langle \vec{S}_1\vec{S}_2\rangle(R,t)$ for the distance $\kf R/\pi=0.51$ and different couplings $J$.
	(b) The reduced correlation function $f_{0.51}(t)$ plotted against the rescaled time $t/t_{0.51}^\mathrm{cor}$.
	NRG parameters: $\lambda=6$, $\mathrm{Ns}=2000$, $N_z=32$, and a TD-NRG damping $ \alpha=0.2$.
}
\label{fig:longtime_R051}
\end{figure}

\begin{figure}[t]
	\centering
  \includegraphics[width=0.5\textwidth]{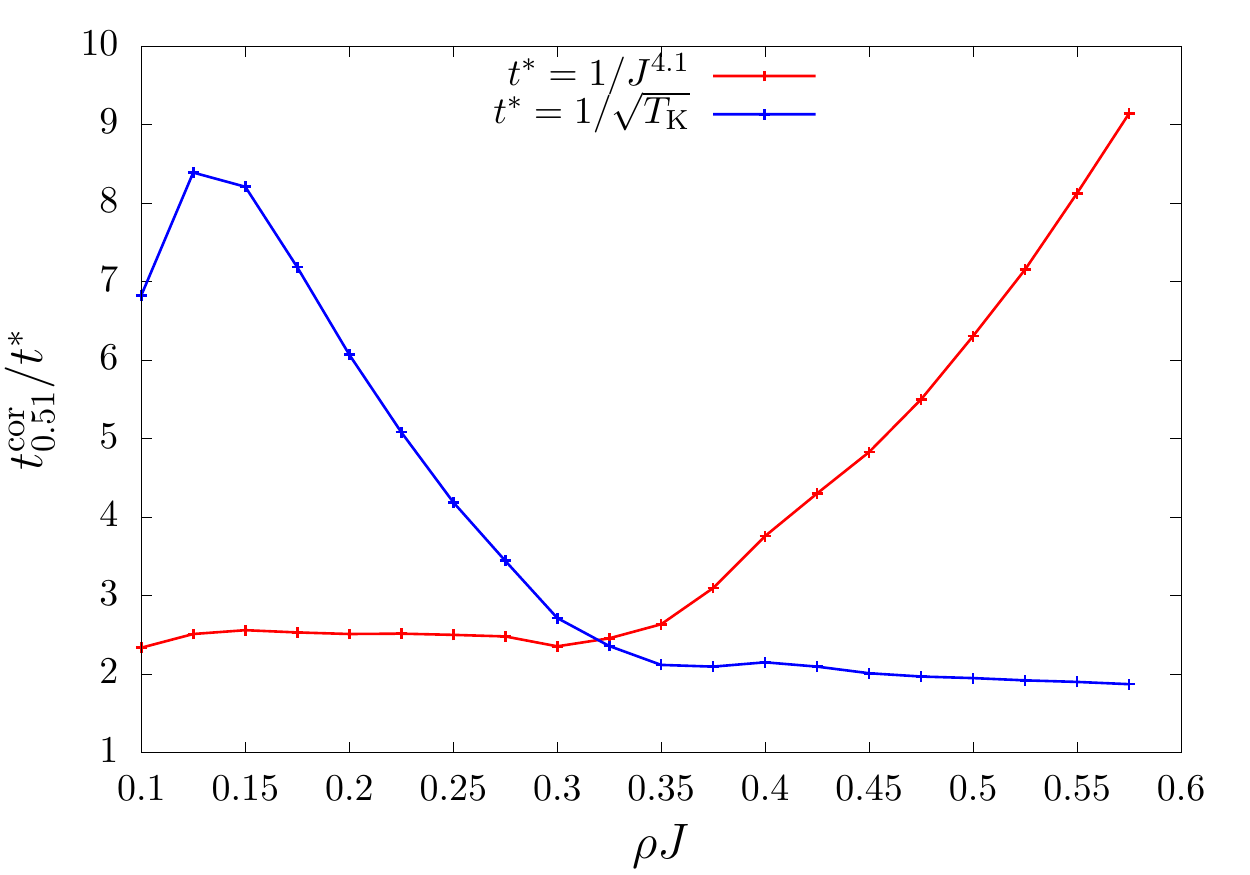}
\caption{(Color online) 
	The rescaled timescales $t_{0.51}^\mathrm{cor} J^{4.13}$ (red line) and $t_{0.51}^\mathrm{cor} \sqrt{T_\mathrm{K}}$ (blue line) plotted against $\rho J$.
}
\label{fig:tStar_Scaling}
\end{figure}

We now turn to the investigation of the long-time behavior for different couplings $J$.
Figure \ref{fig:longtime_R051}(a) depicts the time-dependent correlation function for the distance $\kf R/\pi=0.51$ and different couplings $J$.
The correlation function reaches its long-time value $\langle \vec{S}_1\vec{S}_2\rangle(R,t\to \infty)$
faster the stronger the coupling to the conduction band $J$ is.
Furthermore,  the 
long-time value $|\langle \vec{S}_1\vec{S}_2\rangle(R,t\to \infty)|$ is reduced with increasing $J$,
which coincides with the behavior observed in the equilibrium model (see Fig. \ref{fig:equi:antiferromagnetic}(a)).

In order to identify a coupling dependent timescale on which the correlation function decreases and approaches its long-time value,
we introduce the reduced correlation function
\begin{align}
  f_{0.51}(t)=\frac{\langle \vec{S}_1\vec{S}_2\rangle(\kf R/\pi=0.51,t) - \langle \vec{S}_1\vec{S}_2\rangle_\mathrm{min}}{\langle \vec{S}_1\vec{S}_2\rangle_\mathrm{max} - \langle \vec{S}_1\vec{S}_2\rangle_\mathrm{min}},
\end{align}
where $\langle \vec{S}_1\vec{S}_2\rangle_\mathrm{max}$ is the maximum ferromagnetic value
and $\langle \vec{S}_1\vec{S}_2\rangle_\mathrm{min}$ is the value of the minimum after the decrease
\footnote{For large couplings $\rho J> 0.50$ a second minimum prior to the first one slowly starts to develop the value of which may even become smaller
than the value of the original second minimum for very large couplings $\rho J>0.55$.
In order to achieve comparability with the curves for smaller couplings, we thus use the value of the second minimum as $<\vec{S}_1\vec{S}_2>_{\text{min}}$ for couplings $\rho J>0.55$.}.
We use this function to define the coupling dependent timescale $t_{0.51}^\mathrm{cor}$ by the condition $ f_{0.51}(t_{0.51}^\mathrm{cor})=0.25$.
Figure \ref{fig:longtime_R051}(b) shows the reduced correlation function $f_{0.51}(t)$ plotted 
versus the rescaled time $t/t_{0.51}^\mathrm{cor}$ for different couplings $J$.
We identify two distinct universal behaviors: one for small couplings $\rho J<0.3$ (solid lines) and one for larger couplings $\rho J> 0.45$ (dashes lines).
While for small couplings $J$ the RKKY interaction drives the physics, for larger couplings the Kondo effect becomes dominant. 
This is in accordance with the equilibrium physics discussed before.

For small couplings the inverse timescale $1/t_{0.51}^\mathrm{cor}$ shows a power-law dependence
$1/t_{0.51}^\mathrm{cor}\propto J^{4.1}$, which is very close to $K_\mathrm{RKKY}^2\propto J^4$.
In contrast, for larger couplings we observe an exponential dependency on $J$ which agrees very well with $\sqrt{T_\mathrm{K}}$. 
In order to visualize the two different dependencies of the timescale $t_{0.51}^\mathrm{cor}$,
Fig.\ \ref{fig:tStar_Scaling} shows the rescaled timescale $t_{0.51}^\mathrm{cor}\cdot J^{4.1}$ (red line) and $t_{0.51}^\mathrm{cor} \cdot \sqrt{T_\mathrm{K}}$ (blue line) plotted against $\rho J$.
While for small couplings $t_{0.51}^\mathrm{cor}\cdot J^{4.1}$ is almost constant, it starts to increase for $\rho J>0.3$.
On the other hand, for large couplings $\rho J >0.4$, the curve $t_{0.51}^\mathrm{cor} \cdot \sqrt{T_\mathrm{K}}$ is almost constant.
This quantifies that 
the crossover between an RKKY dominated physics for small $J$ and a Kondo driven physics
for large $J$ is also found in the characteristic timescales of the non-equilibrium dynamics.

\begin{figure}[t]
	\centering
  \flushleft{(a)}
	\includegraphics[width=0.5\textwidth]{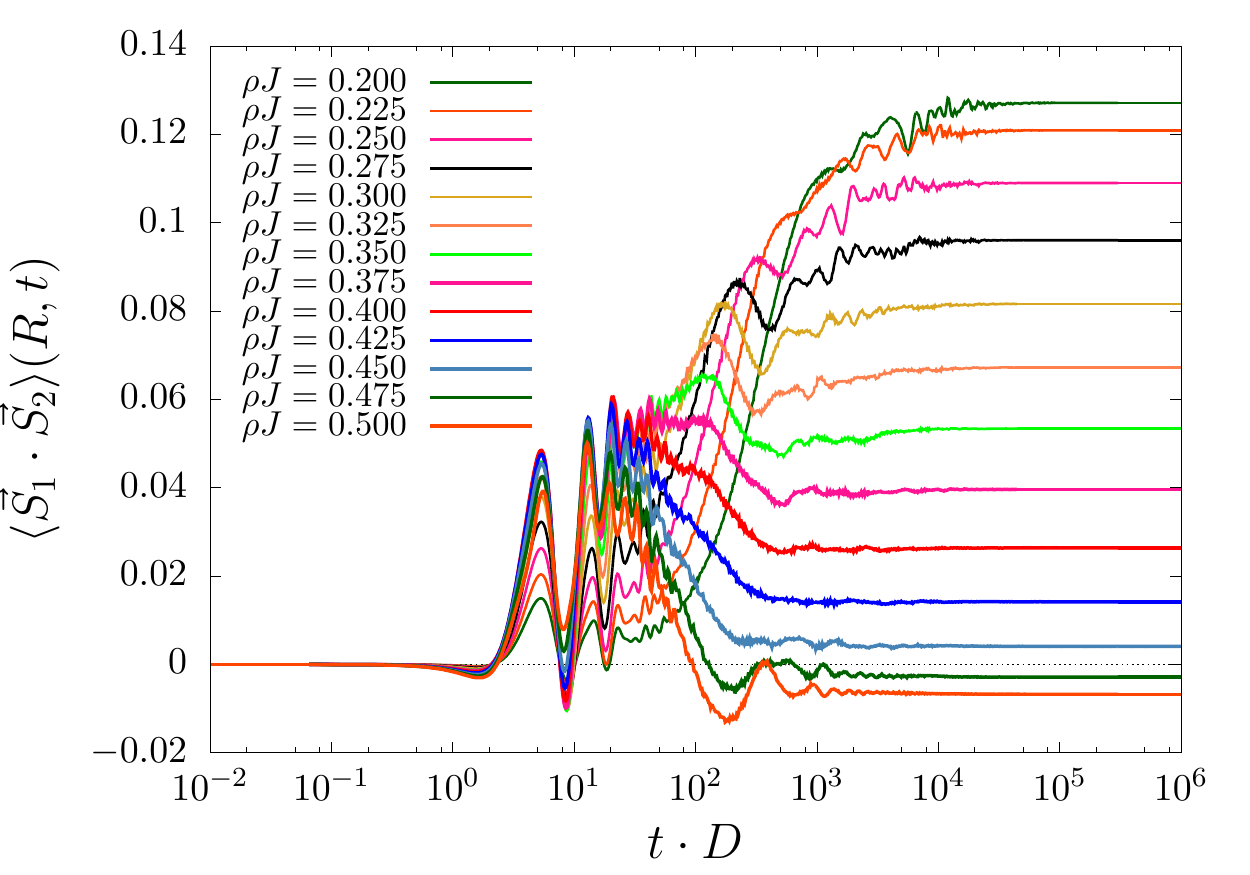}
  \flushleft{(b)}
  \includegraphics[width=0.5\textwidth]{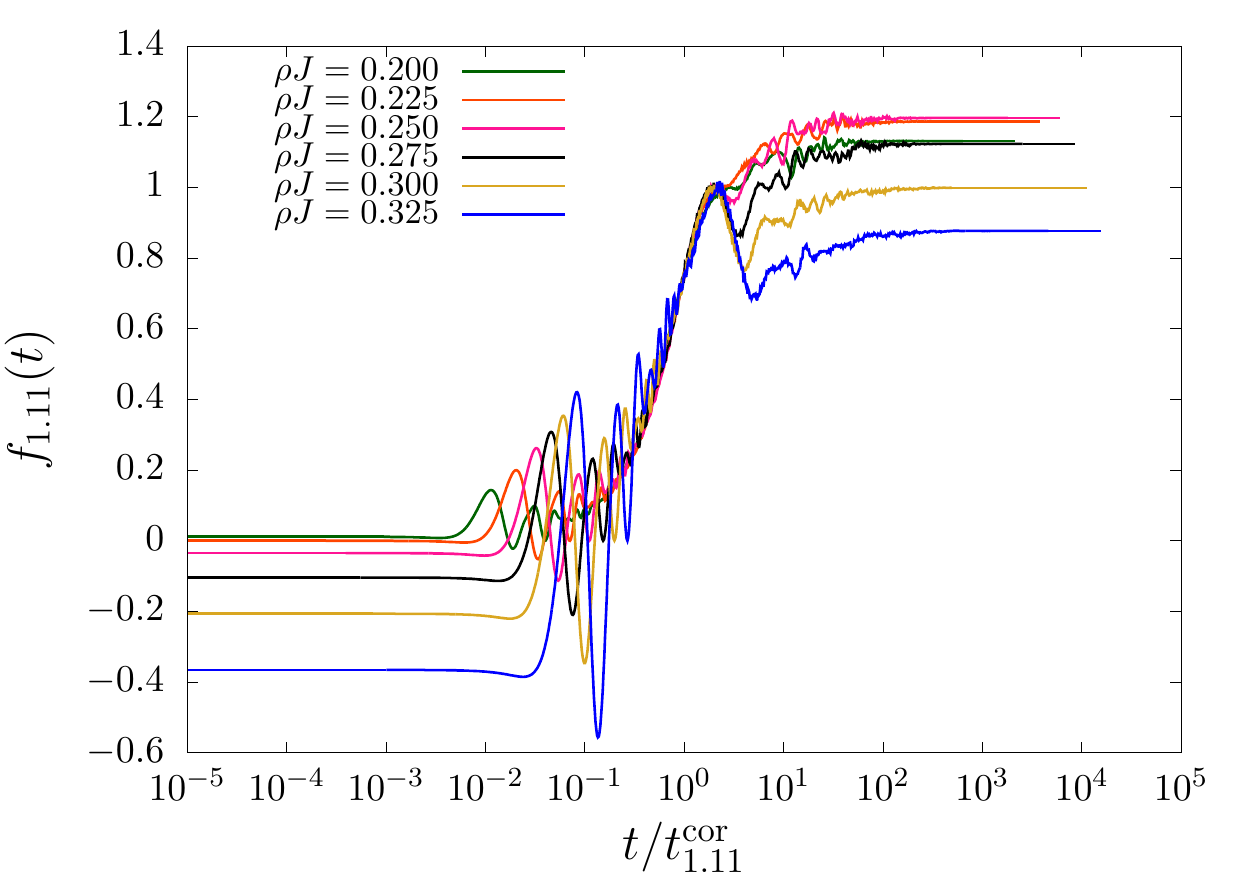}
\caption{(Color online) 
	(a) Time-dependent behavior of the correlation function for the distance $\kf R/\pi=1.11$ and different couplings $J$.
	(b) The reduced correlation function $f_{1.11}(t)$ plotted against the rescaled time $t/t_{1.11}^\mathrm{cor}$.
	NRG parameters: $\lambda=6$, $\mathrm{Ns}=2000$, $N_z=32$, and a TD-NRG damping $ \alpha=0.2$.
}
\label{fig:longtime_R111}
\end{figure}

Figure \ref{fig:longtime_R111} shows the long-time
behavior of the correlation function for different couplings and the distance $\kf R/\pi=1.11$.
Since  the RKKY interaction is ferromagnetic for this distance, the correlation function 
increases after the ferromagnetic wave has passed.
We observe that  the correlation function reaches its long-time value faster
with increasing coupling strength $J$  while the
long-time value $\langle \vec{S}_1\vec{S}_2\rangle(1.11,t\to\infty)$ is reduced.

Interestingly, for large couplings $\rho J>0.3$ the correlation function first increases until its starts to decrease and can even reach an antiferromagnetic long-time value for
couplings $\rho J \geq 0.475$.
This behavior is in accordance with equilibrium results at low temperatures
such that for the distance  $\kf R/\pi=1.11$ and couplings $\rho J \geq 0.475$
we also observe small antiferromagnetic correlation functions in the equilibrium NRG results.
This effect has already been discussed in Sec. \ref{sec:EquilibriumAntiferromagnetic}.

To extract a $J$ dependent timescale, we
again define a reduced correlation function
\begin{align}
  f_{1.11}(t)=\frac{\langle \vec{S}_1\vec{S}_2\rangle(\kf R/\pi=1.11,t) - \langle \vec{S}_1\vec{S}_2\rangle_\mathrm{min}}{\langle \vec{S}_1\vec{S}_2\rangle_\mathrm{max} - \langle \vec{S}_1\vec{S}_2\rangle_\mathrm{min}},
\end{align}
where $\langle \vec{S}_1\vec{S}_2\rangle_\mathrm{min}$ is the value of the second minimum 
of $ \langle \vec{S}_1\vec{S}_2\rangle(\kf R/\pi=1.11,t)$
after the first ferromagnetic peak and 
$\langle \vec{S}_1\vec{S}_2\rangle_\mathrm{max}$ is the value of the maximum of
the same function
directly after the increase  and before the correlation function starts to decrease again.
Here, we modify the definition of the coupling dependent timescale to $f_{1.11}(t_{1.11}^\mathrm{cor})=0.75$.

The reduced correlation function $f_{1.11}(t)$ for small couplings plotted against the rescaled time $t/t_{1.11}^\mathrm{cor}$ is depicted in Fig.\ \ref{fig:longtime_R111}(b).
Due to the rescaling, we find a universal behavior for the increase.
The coupling dependency of the timescale is once again given by $t_{1.11}^\mathrm{cor} \propto J^{-4.1}$.
We can, therefore, conclude that for small couplings $J$ the timescale for the long-time behavior is the same and does not depend on whether the RKKY interaction is ferromagnetic or antiferromagnetic.

The examination of the timescales for larger couplings, however, turns out to be difficult since, as already mentioned above,
the long-time behavior starts to become more complicated than a rather simple increase of the correlation function and instead starts to decrease for long times.

\subsection{Propagation of the correlation}
\label{sec:Nonequilibrium_Propagation}

In this section, we investigate the propagation of correlations through the system.
For that purpose, we combine the real-time dynamics calculations for different but fixed distances
of the two impurities into  two-dimensional plots where the horizontal axis denotes the
dimensionless distance between the two impurities and the vertical axis denotes the time.

\begin{figure}[t]
\centering
 \flushleft{(a)}
 \includegraphics[width=0.5\textwidth]{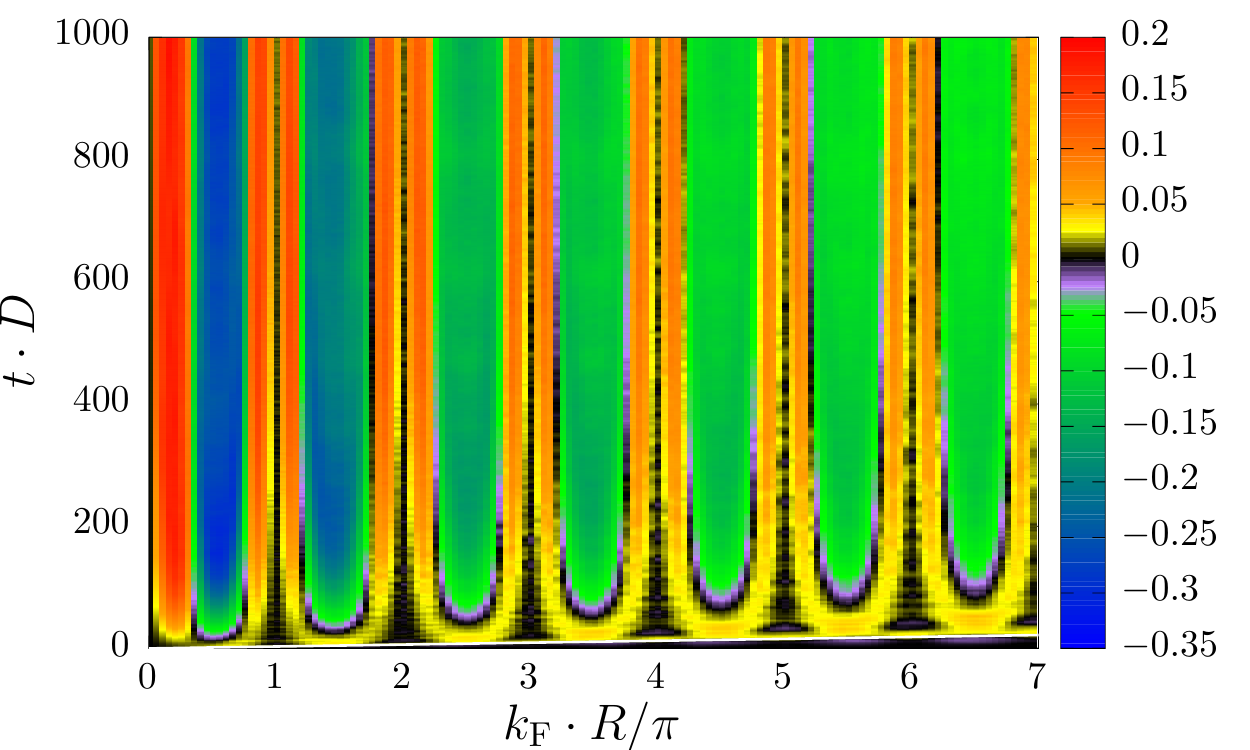}
 \flushleft{(b)}
 \includegraphics[width=0.5\textwidth]{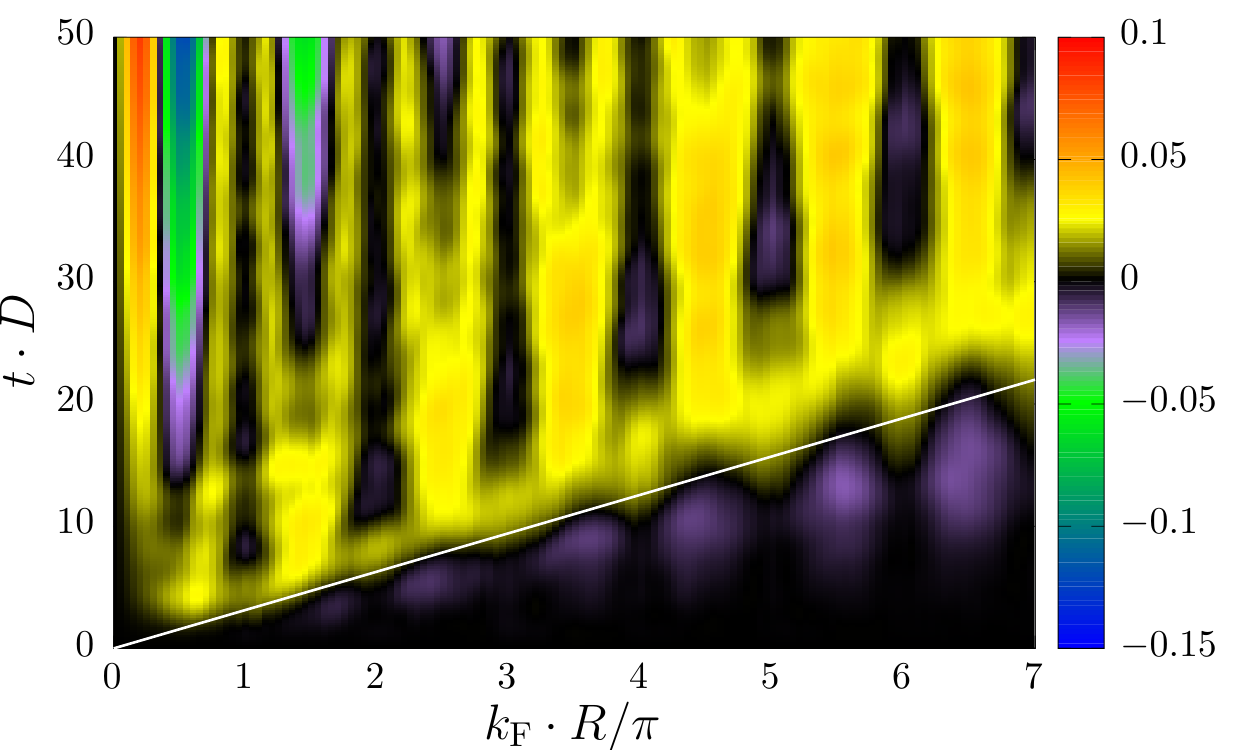}
 \caption{(Color online) 
  (a) Time-dependent correlation function $\langle \vec{S}_1\vec{S}_2\rangle(R,t)$ for $\rho J=0.2$ and a 1D dispersion.
  (b) The short-time behavior in more detail.
  The white line indicates the Fermi velocity $v_\mathrm{F}$.
  Note that for $k_\mathrm{F}R/\pi=n$ the correlation function does not evolve towards its equilibrium value and instead remains almost zero (vertical black lines).
  NRG parameters: $\lambda=3$, $\mathrm{Ns}=1400$, $N_z=4$.
 }
 \label{fig:time:1D}
\end{figure}
For the coupling $\rho J=0.2$ and a 1D dispersion Fig. \ref{fig:time:1D}(a) depicts the correlation function for times up to $tD=1000$ and distances up to $\kf R/\pi=7$.
For long times, $\langle \vec{S}_1\vec{S}_2\rangle(R,t)$ approaches its equilibrium value, and the step like oscillations as found in equilibrium (see Fig.\ \ref{fig:equi:antiferromagnetic})
caused by the RKKY interaction are already clearly visible for times $tD>100$.

In the center of the ferromagnetic correlations at the magic distances $\kf R/\pi=n$,
the black vertical lines indicate that the correlation function remains almost zero.
At these distances the RKKY interaction is maximal ferromagnetic (see Fig. \ref{fig:equi:antiferromagnetic}); however, 
either the even-parity or the odd-parity conduction band decouples from the 
problem.
Therefore, the local impurity parity becomes a conserved quantity which leads to a fixed value for $\langle \vec{S}_1\vec{S}_2\rangle(R,t)$ as already discussed above.

Figure \ref{fig:time:1D}(b) depicts the same data as in Fig. \ref{fig:time:1D}(a) for times up to $tD=50$ 
to illustrate the short-time behavior in more detail.
At $\kf R/\pi =0.5$ a ferromagnetic correlation evolves which then propagates with the Fermi velocity, indicated by the white line, through the conduction band.
Directly in front of the light cone, we observe antiferromagnetic correlations at distances $\kf R/\pi=(n+0.5)$.
Such correlations outside of the light cone were also found for the correlations between the impurity spin and the spin density of the conduction band at distance $R$
and could be traced back to the intrinsic correlations of the Fermi sea \cite{lit:lechtenbergAnders14}.
These correlations are already present before the impurities are coupled to the conduction band and are a property of the Fermi sea.

One can also see that for
the distances $\kf R/\pi=(n+0.5)$ the correlation function at first evolves towards a ferromagnetic value for a relatively long time until it later approaches
its expected antiferromagnetic equilibrium value since the RKKY interactions is antiferromagnetic for these distances.

It becomes apparent that the correlation function remains almost zero 
for distances $\kf R/\pi=n$
after the ferromagnetic correlation wave has passed due to the local parity conservation.
Note that with increasing distance $R$ the frequency of the oscillations in $N^\mathrm{1D}_{e/o}(\epsilon,{R})$ increases and, consequently,
the width of the gap becomes narrower so that the energy scale on which the impurities see the gap decreases with $1/R$.
The decreasing energy scale, on the other hand, leads to a linearly increasing timescale $\propto R$ at which $\langle \vec{S}_1\vec{S}_2\rangle(R,t)$ is fixed.

\begin{figure}[t]
\centering
 \includegraphics[width=0.5\textwidth]{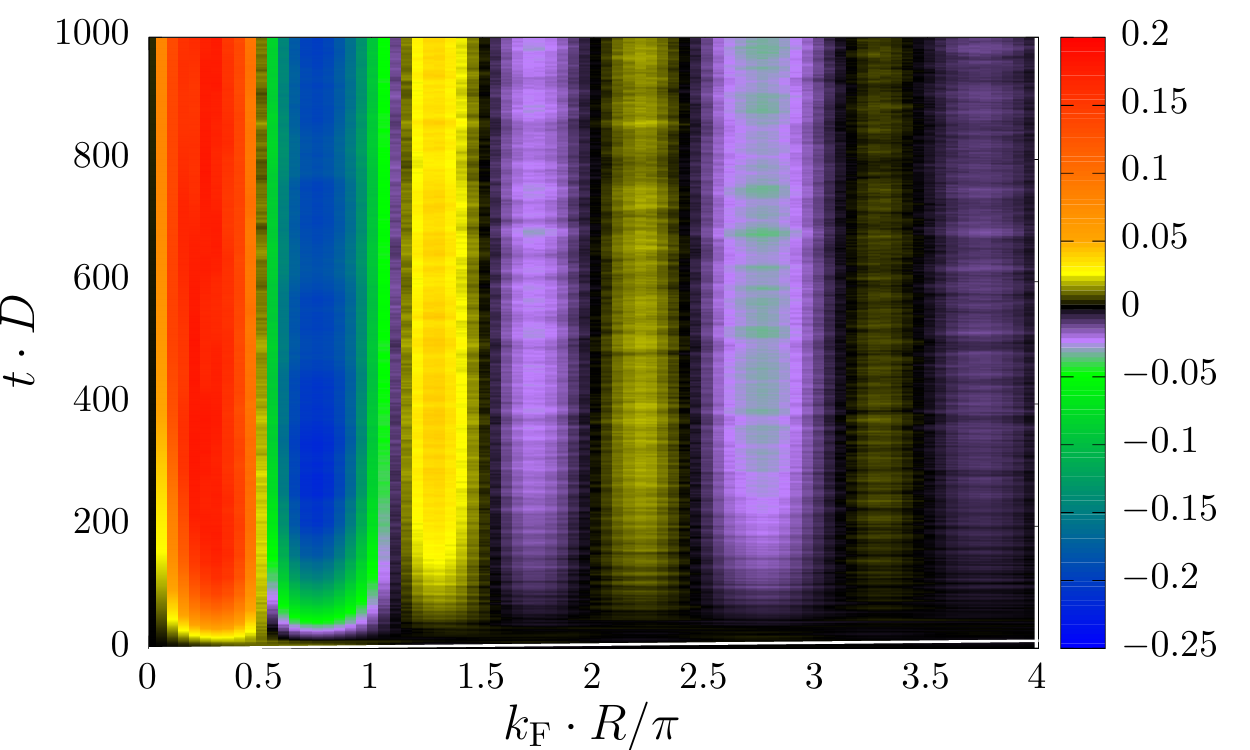}
 \caption{(Color online) 
  Time-dependent correlation function $\langle \vec{S}_1\vec{S}_2\rangle(R,t)$ for a 2D linear dispersion.
  The white line indicates the Fermi velocity $v_\mathrm{F}$.
  NRG parameters: $\lambda=3$, $\mathrm{Ns}=1400$, $N_z=4$.
 }
 \label{fig:time:2D}
\end{figure}

In order to demonstrate
that the local impurity parity conservation is a special feature of certain dispersions,
Fig.\ \ref{fig:time:2D} shows the time-dependent correlation function for a linear dispersion in two dimensions.
The normalization functions are given by 
$N^\mathrm{2D}_{e/o}(\epsilon,{R})=\Gamma_0 [ 1 \pm J_0( \kf R ( 1+\frac{\epsilon}{D}) ) ]$
in this case,
with the zeroth Bessel function $J_0(x)$ \cite{lit:Borda07,lit:lechtenbergAnders14}.
These hybridization functions do not exhibit a gap for any finite distance $R$.
Note that for vanishing distance $R=0$ the odd conduction band always decouples for all dispersions.

In two dimensions we only observe a vanishing correlation function for long times at distances separating the ferromagnetic and antiferromagnetic correlations.
Unlike before, these black vertical lines are simply caused by a vanishing RKKY interaction for these distances.
This is in contrast to the 1D case where $\langle \vec{S}_1\vec{S}_2\rangle(R,t)$ remained zero for distances where the RKKY interaction is maximal ferromagnetic. 
Also note that the correlation function decays faster compared to the 1D case 
for larger distances at large times, which is directly related to
the faster decaying RKKY interaction $\propto 1/R^2$
in comparison to the $\propto 1/R$ decay for a 1D dispersion.

\section{Summary and Outlook}
\label{sec:Summary}

The  equilibrium properties as well as real-time dynamics of 
the spin-correlation function between two localized spins 
at a distance $R$ coupled to one conduction band via a local 
Heisenberg interaction $J$
were investigated using the NRG. Since we did not add
a direct exchange between the  spins, spin-correlations can  only be mediated by the indirect RKKY interaction.

In order to set the stage for the non-equilibrium dynamics after a local interaction quench, we presented
the distance dependent equilibrium spin-correlation function for the TIKM.
There is a competition between
Kondo physics and RKKY mediated singlet formation  \cite{Doniach77,lit:Jones88,lit:Jones1989,lit:Sakai1992_I}
for an AF coupling $J$. For a
FM coupling, the distance dependent spin-correlation function is only weakly coupling dependent due to
reduction of $J$ in the RG. 
For both signs of interactions $J$, the correlation function oscillates with the distance $R$ as expected.
Although the RKKY interactions varies continuously  with the well-established $\cos(2\kf R)$ oscillations in one dimension,
the spin-correlation function $\langle \vec{S}_1\vec{S}_2\rangle(R,t)$ shows a steplike behavior that is
a reminiscent of the zero-temperature level crossing of local singlet-triplet state energies.

For distances $R$ with generically FM RKKY interactions close to its distance dependent
maximum, $\langle \vec{S}_1\vec{S}_2\rangle(R,t)$ clearly reveals the influence of the Kondo screening.
While for $R<\xik$, the correlation function is ferromagnetic as expected, $\langle \vec{S}_1\vec{S}_2\rangle(R,t)$
can change its sign once $R$ exceeds the Kondo correlation length $\xik$.
For $R\to \infty$, two independent Kondo singlets are formed and the spin correlation function vanishes.
At finite distances and $R\gg \xik$, the sign of $\langle \vec{S}_1\vec{S}_2\rangle(R,t)$ depends on the magnitude of
the potential scattering terms. The difference of these terms in the even and odd channel is related to 
a marginal relevant operator  \cite{lit:Affleck95} that generates a small antiferromagnetic interaction responsible for
the sign change.

For distances with purely AF RKKY interactions, at distances $\kf R/\pi=(n+1/2)$, we found universality in $R/\xik$ 
for the amplitude of the correlation function and a $1/R^2$ decay once the distance exceeds $\xik$, which is faster than the $1/R$ decrease of the 1D RKKY interaction:
The Kondo screening of  each impurity spin induces a faster decay of the correlation function.

In the case of ferromagnetic Kondo couplings $J<0$, the amplitude 
remains constant even for $R\to \infty$ since the Kondo effect is absent.
Only finite temperature evokes a power-law decay of the correlation function, which turns into an exponential decay
once the length scale of the finite temperature $\xi_T$ is exceeded.

The non-equilibrium dynamics of the spin-correlation function after a sudden quench
shows distinct behavior for short and for long times as a function of the distance. The short-time
dynamics is governed by the propagation of correlations via the conduction band \cite{lit:lechtenbergAnders14}
with the Fermi velocity: A short ferromagnetic wave is propagating through the system as a consequence
of the total spin conservation since locally anti-ferromagnetic correlations between the local
spin and the local conduction electron spin density are building up. Its magnitude is defined 
by $J^3$, which can be understood
from third-order perturbation theory.

We extracted the characteristic long timescale $t^*$ for a fixed short distance 
reflecting  the different mechanism in the real-time dynamics. While for weak
coupling $J$, the scaling $t^*\propto J^{-4}$ is related to the dominating RKKY interaction,
 $t^*\propto 1/\sqrt{T_K}$ reveals the dominating Kondo effect with increasing local coupling.

The most striking feature is, however, the remarkable non-equilibrium dynamics at the distances 
$k_\mathrm{F}R/\pi=n$. Although the RKKY interaction reaches its periodic maxima,
the correlation function only changes for short times whereas it remains constant for long times.
This effect originates from the
symmetry of the 1D dispersion and is caused by the fact that for these distances
conduction electron states with even-parity ($n=1,3,\dots$) or odd-parity ($n=0,2,\dots$) symmetry decouple from the impurities at low temperatures.
This decoupling also enforces a dynamic local parity conservation for the impurity spins which leads to a conserved 
value of the correlation function for long times.

This effect might be very useful for the implementation of spin qubits since the 
parity symmetry protects the entanglement between both spins and prevents the correlations from decaying to its equilibrium value.
Usually, highly localized electrons in quantum dots are used as qubits since the localization reduces the decoherence facilitated by free-electron motion, but simultaneously increases the
hyperfine interaction strength between the confined electron spin and the surrounding nuclear spins \cite{lit:Hanson2007,lit:Merkulov2002,lit:Loss2004,lit:Loss2008}.
Making use of symmetries such as the parity to retain the entanglement might, therefore, be a way to employ more delocalized electrons and thus decrease the hyperfine interaction.

\begin{acknowledgments}
B.L. thanks the Japan Society for the Promotion of Science (JSPS) and the Alexander von Humboldt Foundation.
	Parts of the computations were performed at the Supercomputer Center, Institute for Solid State Physics, University of Tokyo 
	and the John von Neumann Institute for Computing at the Forschungszentrum Jülich under Project No. HHB00
\end{acknowledgments}

\appendix
\section{Perturbative Approach for the Real-Time Correlation Function} \label{app:perturbation_theory}

In this Appendix we will briefly present a perturbation theory to show
that the lowest non-vanishing contribution to
the real-time dynamics of the correlation function is given by the third order $\propto J^3$.

For this purpose the Hamiltonian is divided into two parts \text{$H=H_0+H_K$} with $H_0=\sum_{\sigma,\vec{k}} \epsilon_{\vec{k}}\, c^\dagger_{\vec{k}\sigma}c^{\phantom{\dagger}}_{\vec{k}\sigma}$, 
the free conduction-band dispersion $\epsilon_{\vec{k}}$, and $H_K= J \sum_i \vec{S}_i \vec{s}_c(r_i)$.
The time-dependent spin-correlation function $\langle \vec{S}_1\vec{S}_2\rangle(t)$ can be written as
\begin{align}
	\langle \vec{S}_1\vec{S}_2\rangle(t)=\textrm{Tr}\left[ \rho^I(t) \vec{S}_1\vec{S}_2\right], 
	\label{eq:app:perturbation_theory:1}
\end{align}
where the index $I$ indicates that the operator is transformed into the interaction picture, which is defined for any operator $A$ as
\begin{align}
	A^I(t)=e^{iH_0t}Ae^{-iH_0t}.
\end{align}
$\vec{S}_1$ and $\vec{S}_1$ remain time independent since they commute with $H_0$.
The von Neumann equation governs the real-time evolution of $\rho^I(t)$,
\begin{align}
	\partial_t \rho^I(t)=i\left[ \rho^I(t), H^I_K(t) \right],
\end{align}
which is integrated to
\begin{align}
	\rho^I(t)=&\rho_0+i\int_0^{t} \left[ \rho_0,H^I_K(t_1) \right] \; dt_1 \nonumber \\
		  & - \int_0^{t}\int_0^{t_1} \left[ \left[\rho^I(t_2),H^I_K(t_2) \right],H^I_K(t_1) \right] \;dt_2\; dt_1,  
	\label{eq:app:perturbation_theory:4}
\end{align}
where we used the initial condition $\rho^I(0)=\rho_0$.
Replacing $\rho^I(t_2)$ by $\rho_0$ in the second integral yields an approximate solution in $O(J^2)$.
Substituting 
\eqref{eq:app:perturbation_theory:4} into \eqref{eq:app:perturbation_theory:1} and performing a cyclically rotation of the operators under the trace, we obtain
\begin{align}
	&\langle \vec{S}_1\vec{S}_2\rangle(t)\approx  \tr{\rho_0\vec{S}_1\vec{S}_2} \nonumber \\
					  &+ i \int_0^t \tr{\rho_0\left[H^I_K(t_1),\vec{S}_1\vec{S}_2 \right]} dt_1  \\
					  &- \int_0^t \int_0^{t_1} \tr{\rho_0\left[H^I_K(t_2),\left[H^I_K(t_1),\vec{S}_1\vec{S}_2\right]\right]} \; dt_2 \; dt_1. \nonumber
\end{align}
This expression contains only expectation values that involve the initial density operator $\rho_0$ in which the impurity spins and the conduction electrons factorize
since in $H_0$ the impurity spins are decoupled from the conduction-band.
In the absence of magnetic fields the first term vanishes, $\tr{\rho_0\vec{S}_1\vec{S}_2}=\langle\vec{S}_1\vec{S}_2\rangle_0=0$, 
where the index denotes that the expectation value is taken with respect to the initial density operator $\rho_0$.

For the integral kernel of the first-order correction we obtain
\begin{align}
	\langle \left[H^I_K(t_1),\vec{S}_1\vec{S}_2 \right] \rangle_0 =& -J \sum_{ijk} \epsilon^{ijk} \left( \langle S_1^k S_2^j \rangle_0 \langle s^i(r_1,t_1) \rangle_0 \right. \nonumber \\
						  &+ \left. \langle S_1^j S_2^k \rangle_0 \langle s^i(r_2,t_1) \rangle_0 \right) =0,
\end{align}
where the upper index indicates the spin component, $\epsilon^{ijk}$ is the Levi-Civita symbol 
and $s^i(r_j,t)$ is the time-dependent spin component of the conduction-band electrons at position $r_j$ in the interaction picture.
Since all occurring expectation values vanish, also the complete first-order contribution in $J$ vanishes.

Calculating the commutator of the second order yields only terms that are proportional to $\langle S^i_1 S_2^j\rangle_0$, $\langle S^i_1 S^j_2 S^k_2 \rangle_0$, or $\langle S^i_1 S^j_1 S^k_2 \rangle_0$.
Since all of these expectation values vanish, the second-order contribution is also zero.

In order to gain a non-vanishing contribution, a finite expectation values is needed which is, e.g.,
given by $\langle S^i_1 S^i_1 S^j_2 S^j_2 \rangle_0$.
Such terms will occur the first time in the third-order contribution $\propto J^3$.
Therefore, the lowest-order contribution to the short-time behavior of the correlation function is given by the third order.
This is in accordance with the TD-NRG results that show a $\propto J^3$ dependence for short times.


\begin{thebibliography}{62}%
\makeatletter
\providecommand \@ifxundefined [1]{%
 \@ifx{#1\undefined}
}%
\providecommand \@ifnum [1]{%
 \ifnum #1\expandafter \@firstoftwo
 \else \expandafter \@secondoftwo
 \fi
}%
\providecommand \@ifx [1]{%
 \ifx #1\expandafter \@firstoftwo
 \else \expandafter \@secondoftwo
 \fi
}%
\providecommand \natexlab [1]{#1}%
\providecommand \enquote  [1]{``#1''}%
\providecommand \bibnamefont  [1]{#1}%
\providecommand \bibfnamefont [1]{#1}%
\providecommand \citenamefont [1]{#1}%
\providecommand \href@noop [0]{\@secondoftwo}%
\providecommand \href [0]{\begingroup \@sanitize@url \@href}%
\providecommand \@href[1]{\@@startlink{#1}\@@href}%
\providecommand \@@href[1]{\endgroup#1\@@endlink}%
\providecommand \@sanitize@url [0]{\catcode `\\12\catcode `\$12\catcode
  `\&12\catcode `\#12\catcode `\^12\catcode `\_12\catcode `\%12\relax}%
\providecommand \@@startlink[1]{}%
\providecommand \@@endlink[0]{}%
\providecommand \url  [0]{\begingroup\@sanitize@url \@url }%
\providecommand \@url [1]{\endgroup\@href {#1}{\urlprefix }}%
\providecommand \urlprefix  [0]{URL }%
\providecommand \Eprint [0]{\href }%
\providecommand \doibase [0]{http://dx.doi.org/}%
\providecommand \selectlanguage [0]{\@gobble}%
\providecommand \bibinfo  [0]{\@secondoftwo}%
\providecommand \bibfield  [0]{\@secondoftwo}%
\providecommand \translation [1]{[#1]}%
\providecommand \BibitemOpen [0]{}%
\providecommand \bibitemStop [0]{}%
\providecommand \bibitemNoStop [0]{.\EOS\space}%
\providecommand \EOS [0]{\spacefactor3000\relax}%
\providecommand \BibitemShut  [1]{\csname bibitem#1\endcsname}%
\let\auto@bib@innerbib\@empty
\bibitem [{\citenamefont {Loss}\ and\ \citenamefont
  {DiVincenzo}(1998)}]{lit:Loss_DiVincenzo1998}%
  \BibitemOpen
  \bibfield  {author} {\bibinfo {author} {\bibfnamefont {D.}~\bibnamefont
  {Loss}}\ and\ \bibinfo {author} {\bibfnamefont {D.~P.}\ \bibnamefont
  {DiVincenzo}},\ }\href {\doibase 10.1103/PhysRevA.57.120} {\bibfield
  {journal} {\bibinfo  {journal} {Phys. Rev. A}\ }\textbf {\bibinfo {volume}
  {57}},\ \bibinfo {pages} {120} (\bibinfo {year} {1998})}\BibitemShut
  {NoStop}%
\bibitem [{\citenamefont {Burkard}\ \emph {et~al.}(1999)\citenamefont
  {Burkard}, \citenamefont {Loss},\ and\ \citenamefont
  {DiVincenzo}}]{lit:Loss_DiVincenzo1999}%
  \BibitemOpen
  \bibfield  {author} {\bibinfo {author} {\bibfnamefont {G.}~\bibnamefont
  {Burkard}}, \bibinfo {author} {\bibfnamefont {D.}~\bibnamefont {Loss}}, \
  and\ \bibinfo {author} {\bibfnamefont {D.~P.}\ \bibnamefont {DiVincenzo}},\
  }\href {\doibase 10.1103/PhysRevB.59.2070} {\bibfield  {journal} {\bibinfo
  {journal} {Phys. Rev. B}\ }\textbf {\bibinfo {volume} {59}},\ \bibinfo
  {pages} {2070} (\bibinfo {year} {1999})}\BibitemShut {NoStop}%
\bibitem [{\citenamefont {Trauzettel}\ \emph {et~al.}(2007)\citenamefont
  {Trauzettel}, \citenamefont {Bulaev}, \citenamefont {Loss},\ and\
  \citenamefont {Burkard}}]{lit:Trauzettel2007}%
  \BibitemOpen
  \bibfield  {author} {\bibinfo {author} {\bibfnamefont {B.}~\bibnamefont
  {Trauzettel}}, \bibinfo {author} {\bibfnamefont {D.~V.}\ \bibnamefont
  {Bulaev}}, \bibinfo {author} {\bibfnamefont {D.}~\bibnamefont {Loss}}, \ and\
  \bibinfo {author} {\bibfnamefont {G.}~\bibnamefont {Burkard}},\ }\href
  {http://dx.doi.org/10.1038/nphys544} {\bibfield  {journal} {\bibinfo
  {journal} {Nature Physics}\ }\textbf {\bibinfo {volume} {3}},\ \bibinfo
  {pages} {192} (\bibinfo {year} {2007})}\BibitemShut {NoStop}%
\bibitem [{\citenamefont {Greilich}\ \emph {et~al.}(2006)\citenamefont
  {Greilich}, \citenamefont {Yakovlev}, \citenamefont {Shabaev}, \citenamefont
  {Efros}, \citenamefont {Yugova}, \citenamefont {Oulton}, \citenamefont
  {Stavarache}, \citenamefont {Reuter}, \citenamefont {Wieck},\ and\
  \citenamefont {Bayer}}]{lit:Greilich2006}%
  \BibitemOpen
  \bibfield  {author} {\bibinfo {author} {\bibfnamefont {A.}~\bibnamefont
  {Greilich}}, \bibinfo {author} {\bibfnamefont {D.~R.}\ \bibnamefont
  {Yakovlev}}, \bibinfo {author} {\bibfnamefont {A.}~\bibnamefont {Shabaev}},
  \bibinfo {author} {\bibfnamefont {A.~L.}\ \bibnamefont {Efros}}, \bibinfo
  {author} {\bibfnamefont {I.~A.}\ \bibnamefont {Yugova}}, \bibinfo {author}
  {\bibfnamefont {R.}~\bibnamefont {Oulton}}, \bibinfo {author} {\bibfnamefont
  {V.}~\bibnamefont {Stavarache}}, \bibinfo {author} {\bibfnamefont
  {D.}~\bibnamefont {Reuter}}, \bibinfo {author} {\bibfnamefont
  {A.}~\bibnamefont {Wieck}}, \ and\ \bibinfo {author} {\bibfnamefont
  {M.}~\bibnamefont {Bayer}},\ }\href {\doibase 10.1126/science.1128215}
  {\bibfield  {journal} {\bibinfo  {journal} {Science}\ }\textbf {\bibinfo
  {volume} {313}},\ \bibinfo {pages} {341} (\bibinfo {year} {2006})},\ \Eprint
  {http://arxiv.org/abs/http://science.sciencemag.org/content/313/5785/341.full.pdf}
  {http://science.sciencemag.org/content/313/5785/341.full.pdf} \BibitemShut
  {NoStop}%
\bibitem [{\citenamefont {Glazov}(2013)}]{lit:Glazov2013}%
  \BibitemOpen
  \bibfield  {author} {\bibinfo {author} {\bibfnamefont {M.~M.}\ \bibnamefont
  {Glazov}},\ }\href {\doibase 10.1063/1.4795515} {\bibfield  {journal}
  {\bibinfo  {journal} {Journal of Applied Physics}\ }\textbf {\bibinfo
  {volume} {113}},\ \bibinfo {pages} {136503} (\bibinfo {year} {2013})},\
  \Eprint {http://arxiv.org/abs/https://doi.org/10.1063/1.4795515}
  {https://doi.org/10.1063/1.4795515} \BibitemShut {NoStop}%
\bibitem [{\citenamefont {\ifmmode \check{Z}\else
  \v{Z}\fi{}uti\ifmmode~\acute{c}\else \'{c}\fi{}}\ \emph
  {et~al.}(2004)\citenamefont {\ifmmode \check{Z}\else
  \v{Z}\fi{}uti\ifmmode~\acute{c}\else \'{c}\fi{}}, \citenamefont {Fabian},\
  and\ \citenamefont {Das~Sarma}}]{lit:RevModPhys.76.323}%
  \BibitemOpen
  \bibfield  {author} {\bibinfo {author} {\bibfnamefont {I.}~\bibnamefont
  {\ifmmode \check{Z}\else \v{Z}\fi{}uti\ifmmode~\acute{c}\else \'{c}\fi{}}},
  \bibinfo {author} {\bibfnamefont {J.}~\bibnamefont {Fabian}}, \ and\ \bibinfo
  {author} {\bibfnamefont {S.}~\bibnamefont {Das~Sarma}},\ }\href {\doibase
  10.1103/RevModPhys.76.323} {\bibfield  {journal} {\bibinfo  {journal} {Rev.
  Mod. Phys.}\ }\textbf {\bibinfo {volume} {76}},\ \bibinfo {pages} {323}
  (\bibinfo {year} {2004})}\BibitemShut {NoStop}%
\bibitem [{\citenamefont {Misiorny}\ \emph {et~al.}(2013)\citenamefont
  {Misiorny}, \citenamefont {Hell},\ and\ \citenamefont
  {Wegewijs}}]{lit:Spintronic_anisotropy_Misiorny2013}%
  \BibitemOpen
  \bibfield  {author} {\bibinfo {author} {\bibfnamefont {M.}~\bibnamefont
  {Misiorny}}, \bibinfo {author} {\bibfnamefont {M.}~\bibnamefont {Hell}}, \
  and\ \bibinfo {author} {\bibfnamefont {M.~R.}\ \bibnamefont {Wegewijs}},\
  }\href {http://dx.doi.org/10.1038/nphys2766} {\bibfield  {journal} {\bibinfo
  {journal} {Nat Phys}\ }\textbf {\bibinfo {volume} {9}},\ \bibinfo {pages}
  {801} (\bibinfo {year} {2013})}\BibitemShut {NoStop}%
\bibitem [{\citenamefont {Han}\ \emph {et~al.}(2014)\citenamefont {Han},
  \citenamefont {Kawakami}, \citenamefont {Gmitra},\ and\ \citenamefont
  {Fabian}}]{lit:Graphene_spintronics_Han2014}%
  \BibitemOpen
  \bibfield  {author} {\bibinfo {author} {\bibfnamefont {W.}~\bibnamefont
  {Han}}, \bibinfo {author} {\bibfnamefont {R.~K.}\ \bibnamefont {Kawakami}},
  \bibinfo {author} {\bibfnamefont {M.}~\bibnamefont {Gmitra}}, \ and\ \bibinfo
  {author} {\bibfnamefont {J.}~\bibnamefont {Fabian}},\ }\href
  {http://dx.doi.org/10.1038/nnano.2014.214} {\bibfield  {journal} {\bibinfo
  {journal} {Nat Nano}\ }\textbf {\bibinfo {volume} {9}},\ \bibinfo {pages}
  {794} (\bibinfo {year} {2014})}\BibitemShut {NoStop}%
\bibitem [{\citenamefont {Johll}\ \emph {et~al.}(2014)\citenamefont {Johll},
  \citenamefont {Lee}, \citenamefont {Ng}, \citenamefont {Kang},\ and\
  \citenamefont {Tok}}]{lit:Metals_on_silicon_Johll2014}%
  \BibitemOpen
  \bibfield  {author} {\bibinfo {author} {\bibfnamefont {H.}~\bibnamefont
  {Johll}}, \bibinfo {author} {\bibfnamefont {M.~D.~K.}\ \bibnamefont {Lee}},
  \bibinfo {author} {\bibfnamefont {S.~P.~N.}\ \bibnamefont {Ng}}, \bibinfo
  {author} {\bibfnamefont {H.~C.}\ \bibnamefont {Kang}}, \ and\ \bibinfo
  {author} {\bibfnamefont {E.~S.}\ \bibnamefont {Tok}},\ }\href
  {http://dx.doi.org/10.1038/srep07594} {\bibfield  {journal} {\bibinfo
  {journal} {Scientific Reports}\ }\textbf {\bibinfo {volume} {4}},\ \bibinfo
  {pages} {7594} (\bibinfo {year} {2014})}\BibitemShut {NoStop}%
\bibitem [{\citenamefont {Yazyev}\ and\ \citenamefont
  {Helm}(2007)}]{lit:Defect_graphene_Yazyev2007}%
  \BibitemOpen
  \bibfield  {author} {\bibinfo {author} {\bibfnamefont {O.~V.}\ \bibnamefont
  {Yazyev}}\ and\ \bibinfo {author} {\bibfnamefont {L.}~\bibnamefont {Helm}},\
  }\href {\doibase 10.1103/PhysRevB.75.125408} {\bibfield  {journal} {\bibinfo
  {journal} {Phys. Rev. B}\ }\textbf {\bibinfo {volume} {75}},\ \bibinfo
  {pages} {125408} (\bibinfo {year} {2007})}\BibitemShut {NoStop}%
\bibitem [{\citenamefont {Bork}\ \emph {et~al.}(2011)\citenamefont {Bork},
  \citenamefont {Zhang}, \citenamefont {Diekhoner}, \citenamefont {Borda},
  \citenamefont {Simon}, \citenamefont {Kroha}, \citenamefont {Wahl},\ and\
  \citenamefont {Kern}}]{lit:Bork_Kroha2011}%
  \BibitemOpen
  \bibfield  {author} {\bibinfo {author} {\bibfnamefont {J.}~\bibnamefont
  {Bork}}, \bibinfo {author} {\bibfnamefont {Y.-h.}\ \bibnamefont {Zhang}},
  \bibinfo {author} {\bibfnamefont {L.}~\bibnamefont {Diekhoner}}, \bibinfo
  {author} {\bibfnamefont {L.}~\bibnamefont {Borda}}, \bibinfo {author}
  {\bibfnamefont {P.}~\bibnamefont {Simon}}, \bibinfo {author} {\bibfnamefont
  {J.}~\bibnamefont {Kroha}}, \bibinfo {author} {\bibfnamefont
  {P.}~\bibnamefont {Wahl}}, \ and\ \bibinfo {author} {\bibfnamefont
  {K.}~\bibnamefont {Kern}},\ }\href {http://dx.doi.org/10.1038/nphys2076}
  {\bibfield  {journal} {\bibinfo  {journal} {Nat Phys}\ }\textbf {\bibinfo
  {volume} {7}},\ \bibinfo {pages} {901} (\bibinfo {year} {2011})}\BibitemShut
  {NoStop}%
\bibitem [{\citenamefont {Esat}\ \emph {et~al.}(2016)\citenamefont {Esat},
  \citenamefont {Lechtenberg}, \citenamefont {Deilmann}, \citenamefont
  {Wagner}, \citenamefont {Kruger}, \citenamefont {Temirov}, \citenamefont
  {Rohlfing}, \citenamefont {Anders},\ and\ \citenamefont
  {Tautz}}]{lit:lechtenberg_2016_dimer}%
  \BibitemOpen
  \bibfield  {author} {\bibinfo {author} {\bibfnamefont {T.}~\bibnamefont
  {Esat}}, \bibinfo {author} {\bibfnamefont {B.}~\bibnamefont {Lechtenberg}},
  \bibinfo {author} {\bibfnamefont {T.}~\bibnamefont {Deilmann}}, \bibinfo
  {author} {\bibfnamefont {C.}~\bibnamefont {Wagner}}, \bibinfo {author}
  {\bibfnamefont {P.}~\bibnamefont {Kruger}}, \bibinfo {author} {\bibfnamefont
  {R.}~\bibnamefont {Temirov}}, \bibinfo {author} {\bibfnamefont
  {M.}~\bibnamefont {Rohlfing}}, \bibinfo {author} {\bibfnamefont {F.~B.}\
  \bibnamefont {Anders}}, \ and\ \bibinfo {author} {\bibfnamefont {F.~S.}\
  \bibnamefont {Tautz}},\ }\href {http://dx.doi.org/10.1038/nphys3737}
  {\bibfield  {journal} {\bibinfo  {journal} {Nat Phys}\ }\textbf {\bibinfo
  {volume} {12}},\ \bibinfo {pages} {867} (\bibinfo {year} {2016})}\BibitemShut
  {NoStop}%
\bibitem [{\citenamefont {Atodiresei}\ \emph {et~al.}(2010)\citenamefont
  {Atodiresei}, \citenamefont {Brede}, \citenamefont
  {Lazi\ifmmode~\acute{c}\else \'{c}\fi{}}, \citenamefont {Caciuc},
  \citenamefont {Hoffmann}, \citenamefont {Wiesendanger},\ and\ \citenamefont
  {Bl\"ugel}}]{lit:spinfilter2010}%
  \BibitemOpen
  \bibfield  {author} {\bibinfo {author} {\bibfnamefont {N.}~\bibnamefont
  {Atodiresei}}, \bibinfo {author} {\bibfnamefont {J.}~\bibnamefont {Brede}},
  \bibinfo {author} {\bibfnamefont {P.}~\bibnamefont
  {Lazi\ifmmode~\acute{c}\else \'{c}\fi{}}}, \bibinfo {author} {\bibfnamefont
  {V.}~\bibnamefont {Caciuc}}, \bibinfo {author} {\bibfnamefont
  {G.}~\bibnamefont {Hoffmann}}, \bibinfo {author} {\bibfnamefont
  {R.}~\bibnamefont {Wiesendanger}}, \ and\ \bibinfo {author} {\bibfnamefont
  {S.}~\bibnamefont {Bl\"ugel}},\ }\href {\doibase
  10.1103/PhysRevLett.105.066601} {\bibfield  {journal} {\bibinfo  {journal}
  {Phys. Rev. Lett.}\ }\textbf {\bibinfo {volume} {105}},\ \bibinfo {pages}
  {066601} (\bibinfo {year} {2010})}\BibitemShut {NoStop}%
\bibitem [{\citenamefont {Bogani}\ and\ \citenamefont
  {Wernsdorfer}(2008)}]{lit:spintronic_review_Bogani2008}%
  \BibitemOpen
  \bibfield  {author} {\bibinfo {author} {\bibfnamefont {L.}~\bibnamefont
  {Bogani}}\ and\ \bibinfo {author} {\bibfnamefont {W.}~\bibnamefont
  {Wernsdorfer}},\ }\href {http://dx.doi.org/10.1038/nmat2133} {\bibfield
  {journal} {\bibinfo  {journal} {Nat Mater}\ }\textbf {\bibinfo {volume}
  {7}},\ \bibinfo {pages} {179} (\bibinfo {year} {2008})}\BibitemShut {NoStop}%
\bibitem [{\citenamefont
  {Sanvito}(2011)}]{lit:Chemistry_Review_spintronics_Sanvito2011}%
  \BibitemOpen
  \bibfield  {author} {\bibinfo {author} {\bibfnamefont {S.}~\bibnamefont
  {Sanvito}},\ }\href {\doibase 10.1039/C1CS15047B} {\bibfield  {journal}
  {\bibinfo  {journal} {Chem. Soc. Rev.}\ }\textbf {\bibinfo {volume} {40}},\
  \bibinfo {pages} {3336} (\bibinfo {year} {2011})}\BibitemShut {NoStop}%
\bibitem [{\citenamefont {Naber}\ \emph {et~al.}(2007)\citenamefont {Naber},
  \citenamefont {Faez},\ and\ \citenamefont {van~der
  Wiel}}]{lit:Review_Naber2007}%
  \BibitemOpen
  \bibfield  {author} {\bibinfo {author} {\bibfnamefont {W.~J.~M.}\
  \bibnamefont {Naber}}, \bibinfo {author} {\bibfnamefont {S.}~\bibnamefont
  {Faez}}, \ and\ \bibinfo {author} {\bibfnamefont {W.~G.}\ \bibnamefont
  {van~der Wiel}},\ }\href {http://stacks.iop.org/0022-3727/40/i=12/a=R01}
  {\bibfield  {journal} {\bibinfo  {journal} {Journal of Physics D: Applied
  Physics}\ }\textbf {\bibinfo {volume} {40}},\ \bibinfo {pages} {R205}
  (\bibinfo {year} {2007})}\BibitemShut {NoStop}%
\bibitem [{\citenamefont {Dediu}\ \emph {et~al.}(2009)\citenamefont {Dediu},
  \citenamefont {Hueso}, \citenamefont {Bergenti},\ and\ \citenamefont
  {Taliani}}]{lit:Organic_Dediu2009}%
  \BibitemOpen
  \bibfield  {author} {\bibinfo {author} {\bibfnamefont {V.~A.}\ \bibnamefont
  {Dediu}}, \bibinfo {author} {\bibfnamefont {L.~E.}\ \bibnamefont {Hueso}},
  \bibinfo {author} {\bibfnamefont {I.}~\bibnamefont {Bergenti}}, \ and\
  \bibinfo {author} {\bibfnamefont {C.}~\bibnamefont {Taliani}},\ }\href
  {http://dx.doi.org/10.1038/nmat2510} {\bibfield  {journal} {\bibinfo
  {journal} {Nat Mater}\ }\textbf {\bibinfo {volume} {8}},\ \bibinfo {pages}
  {707} (\bibinfo {year} {2009})}\BibitemShut {NoStop}%
\bibitem [{\citenamefont {Drew}\ \emph {et~al.}(2009)\citenamefont {Drew},
  \citenamefont {Hoppler}, \citenamefont {Schulz}, \citenamefont {Pratt},
  \citenamefont {Desai}, \citenamefont {Shakya}, \citenamefont {Kreouzis},
  \citenamefont {Gillin}, \citenamefont {Suter}, \citenamefont {Morley},
  \citenamefont {Malik}, \citenamefont {Dubroka}, \citenamefont {Kim},
  \citenamefont {Bouyanfif}, \citenamefont {Bourqui}, \citenamefont {Bernhard},
  \citenamefont {Scheuermann}, \citenamefont {Nieuwenhuys}, \citenamefont
  {Prokscha},\ and\ \citenamefont {Morenzoni}}]{lit:spin_diffusion_Drew2009}%
  \BibitemOpen
  \bibfield  {author} {\bibinfo {author} {\bibfnamefont {A.~J.}\ \bibnamefont
  {Drew}}, \bibinfo {author} {\bibfnamefont {J.}~\bibnamefont {Hoppler}},
  \bibinfo {author} {\bibfnamefont {L.}~\bibnamefont {Schulz}}, \bibinfo
  {author} {\bibfnamefont {F.~L.}\ \bibnamefont {Pratt}}, \bibinfo {author}
  {\bibfnamefont {P.}~\bibnamefont {Desai}}, \bibinfo {author} {\bibfnamefont
  {P.}~\bibnamefont {Shakya}}, \bibinfo {author} {\bibfnamefont
  {T.}~\bibnamefont {Kreouzis}}, \bibinfo {author} {\bibfnamefont {W.~P.}\
  \bibnamefont {Gillin}}, \bibinfo {author} {\bibfnamefont {A.}~\bibnamefont
  {Suter}}, \bibinfo {author} {\bibfnamefont {N.~A.}\ \bibnamefont {Morley}},
  \bibinfo {author} {\bibfnamefont {V.~K.}\ \bibnamefont {Malik}}, \bibinfo
  {author} {\bibfnamefont {A.}~\bibnamefont {Dubroka}}, \bibinfo {author}
  {\bibfnamefont {K.~W.}\ \bibnamefont {Kim}}, \bibinfo {author} {\bibfnamefont
  {H.}~\bibnamefont {Bouyanfif}}, \bibinfo {author} {\bibfnamefont
  {F.}~\bibnamefont {Bourqui}}, \bibinfo {author} {\bibfnamefont
  {C.}~\bibnamefont {Bernhard}}, \bibinfo {author} {\bibfnamefont
  {R.}~\bibnamefont {Scheuermann}}, \bibinfo {author} {\bibfnamefont {G.~J.}\
  \bibnamefont {Nieuwenhuys}}, \bibinfo {author} {\bibfnamefont
  {T.}~\bibnamefont {Prokscha}}, \ and\ \bibinfo {author} {\bibfnamefont
  {E.}~\bibnamefont {Morenzoni}},\ }\href {http://dx.doi.org/10.1038/nmat2333}
  {\bibfield  {journal} {\bibinfo  {journal} {Nat Mater}\ }\textbf {\bibinfo
  {volume} {8}},\ \bibinfo {pages} {109} (\bibinfo {year} {2009})}\BibitemShut
  {NoStop}%
\bibitem [{\citenamefont {Spinelli}\ \emph {et~al.}(2015)\citenamefont
  {Spinelli}, \citenamefont {Gerrits}, \citenamefont {Toskovic}, \citenamefont
  {Bryant}, \citenamefont {Ternes},\ and\ \citenamefont
  {Otte}}]{lit:Spinelli2015}%
  \BibitemOpen
  \bibfield  {author} {\bibinfo {author} {\bibfnamefont {A.}~\bibnamefont
  {Spinelli}}, \bibinfo {author} {\bibfnamefont {M.}~\bibnamefont {Gerrits}},
  \bibinfo {author} {\bibfnamefont {R.}~\bibnamefont {Toskovic}}, \bibinfo
  {author} {\bibfnamefont {B.}~\bibnamefont {Bryant}}, \bibinfo {author}
  {\bibfnamefont {M.}~\bibnamefont {Ternes}}, \ and\ \bibinfo {author}
  {\bibfnamefont {A.~F.}\ \bibnamefont {Otte}},\ }\href
  {http://dx.doi.org/10.1038/ncomms10046} {\bibfield  {journal} {\bibinfo
  {journal} {Nature Communications}\ }\textbf {\bibinfo {volume} {6}},\
  \bibinfo {pages} {10046} (\bibinfo {year} {2015})}\BibitemShut {NoStop}%
\bibitem [{\citenamefont {Jones}\ \emph {et~al.}(1988)\citenamefont {Jones},
  \citenamefont {Varma},\ and\ \citenamefont {Wilkins}}]{lit:Jones88}%
  \BibitemOpen
  \bibfield  {author} {\bibinfo {author} {\bibfnamefont {B.~A.}\ \bibnamefont
  {Jones}}, \bibinfo {author} {\bibfnamefont {C.~M.}\ \bibnamefont {Varma}}, \
  and\ \bibinfo {author} {\bibfnamefont {J.~W.}\ \bibnamefont {Wilkins}},\
  }\href {\doibase 10.1103/PhysRevLett.61.125} {\bibfield  {journal} {\bibinfo
  {journal} {Phys. Rev. Lett.}\ }\textbf {\bibinfo {volume} {61}},\ \bibinfo
  {pages} {125} (\bibinfo {year} {1988})}\BibitemShut {NoStop}%
\bibitem [{\citenamefont {Jones}\ and\ \citenamefont
  {Varma}(1989)}]{lit:Jones1989}%
  \BibitemOpen
  \bibfield  {author} {\bibinfo {author} {\bibfnamefont {B.~A.}\ \bibnamefont
  {Jones}}\ and\ \bibinfo {author} {\bibfnamefont {C.~M.}\ \bibnamefont
  {Varma}},\ }\href {\doibase 10.1103/PhysRevB.40.324} {\bibfield  {journal}
  {\bibinfo  {journal} {Phys. Rev. B}\ }\textbf {\bibinfo {volume} {40}},\
  \bibinfo {pages} {324} (\bibinfo {year} {1989})}\BibitemShut {NoStop}%
\bibitem [{\citenamefont {Fye}\ and\ \citenamefont
  {Hirsch}(1989)}]{lit:Fye1989}%
  \BibitemOpen
  \bibfield  {author} {\bibinfo {author} {\bibfnamefont {R.~M.}\ \bibnamefont
  {Fye}}\ and\ \bibinfo {author} {\bibfnamefont {J.~E.}\ \bibnamefont
  {Hirsch}},\ }\href {\doibase 10.1103/PhysRevB.40.4780} {\bibfield  {journal}
  {\bibinfo  {journal} {Phys. Rev. B}\ }\textbf {\bibinfo {volume} {40}},\
  \bibinfo {pages} {4780} (\bibinfo {year} {1989})}\BibitemShut {NoStop}%
\bibitem [{\citenamefont {Fye}(1994)}]{lit:Fye1994}%
  \BibitemOpen
  \bibfield  {author} {\bibinfo {author} {\bibfnamefont {R.~M.}\ \bibnamefont
  {Fye}},\ }\href {\doibase 10.1103/PhysRevLett.72.916} {\bibfield  {journal}
  {\bibinfo  {journal} {Phys. Rev. Lett.}\ }\textbf {\bibinfo {volume} {72}},\
  \bibinfo {pages} {916} (\bibinfo {year} {1994})}\BibitemShut {NoStop}%
\bibitem [{\citenamefont {Affleck}\ \emph {et~al.}(1995)\citenamefont
  {Affleck}, \citenamefont {Ludwig},\ and\ \citenamefont
  {Jones}}]{lit:Affleck95}%
  \BibitemOpen
  \bibfield  {author} {\bibinfo {author} {\bibfnamefont {I.}~\bibnamefont
  {Affleck}}, \bibinfo {author} {\bibfnamefont {A.~W.~W.}\ \bibnamefont
  {Ludwig}}, \ and\ \bibinfo {author} {\bibfnamefont {B.~A.}\ \bibnamefont
  {Jones}},\ }\href {\doibase 10.1103/PhysRevB.52.9528} {\bibfield  {journal}
  {\bibinfo  {journal} {Phys. Rev. B}\ }\textbf {\bibinfo {volume} {52}},\
  \bibinfo {pages} {9528} (\bibinfo {year} {1995})}\BibitemShut {NoStop}%
\bibitem [{\citenamefont {Ruderman}\ and\ \citenamefont
  {Kittel}(1954)}]{lit:RudermanKittel1954}%
  \BibitemOpen
  \bibfield  {author} {\bibinfo {author} {\bibfnamefont {M.~A.}\ \bibnamefont
  {Ruderman}}\ and\ \bibinfo {author} {\bibfnamefont {C.}~\bibnamefont
  {Kittel}},\ }\href {\doibase 10.1103/PhysRev.96.99} {\bibfield  {journal}
  {\bibinfo  {journal} {Phys. Rev.}\ }\textbf {\bibinfo {volume} {96}},\
  \bibinfo {pages} {99} (\bibinfo {year} {1954})}\BibitemShut {NoStop}%
\bibitem [{\citenamefont {{Kasuya}}(1956)}]{lit:Kasuya1956}%
  \BibitemOpen
  \bibfield  {author} {\bibinfo {author} {\bibfnamefont {T.}~\bibnamefont
  {{Kasuya}}},\ }\href {\doibase 10.1143/PTP.16.45} {\bibfield  {journal}
  {\bibinfo  {journal} {Progress of Theoretical Physics}\ }\textbf {\bibinfo
  {volume} {16}},\ \bibinfo {pages} {45} (\bibinfo {year} {1956})}\BibitemShut
  {NoStop}%
\bibitem [{\citenamefont {Yosida}(1957)}]{lit:Yosida1957}%
  \BibitemOpen
  \bibfield  {author} {\bibinfo {author} {\bibfnamefont {K.}~\bibnamefont
  {Yosida}},\ }\href {\doibase 10.1103/PhysRev.106.893} {\bibfield  {journal}
  {\bibinfo  {journal} {Phys. Rev.}\ }\textbf {\bibinfo {volume} {106}},\
  \bibinfo {pages} {893} (\bibinfo {year} {1957})}\BibitemShut {NoStop}%
\bibitem [{\citenamefont {Doniach}(1977)}]{Doniach77}%
  \BibitemOpen
  \bibfield  {author} {\bibinfo {author} {\bibfnamefont {S.}~\bibnamefont
  {Doniach}},\ }\href@noop {} {\bibfield  {journal} {\bibinfo  {journal}
  {Physica B}\ }\textbf {\bibinfo {volume} {91}},\ \bibinfo {pages} {231}
  (\bibinfo {year} {1977})}\BibitemShut {NoStop}%
\bibitem [{\citenamefont {Sakai}\ and\ \citenamefont
  {Shimizu}(1992)}]{lit:Sakai1992_I}%
  \BibitemOpen
  \bibfield  {author} {\bibinfo {author} {\bibfnamefont {O.}~\bibnamefont
  {Sakai}}\ and\ \bibinfo {author} {\bibfnamefont {Y.}~\bibnamefont
  {Shimizu}},\ }\href {\doibase 10.1143/JPSJ.61.2333} {\bibfield  {journal}
  {\bibinfo  {journal} {Journal of the Physical Society of Japan}\ }\textbf
  {\bibinfo {volume} {61}},\ \bibinfo {pages} {2333} (\bibinfo {year}
  {1992})}\BibitemShut {NoStop}%
\bibitem [{\citenamefont {Silva}\ \emph {et~al.}(1996)\citenamefont {Silva},
  \citenamefont {Lima}, \citenamefont {Oliveira}, \citenamefont {Mello},
  \citenamefont {Oliveira},\ and\ \citenamefont {Wilkins}}]{lit:Silva1995}%
  \BibitemOpen
  \bibfield  {author} {\bibinfo {author} {\bibfnamefont {J.~B.}\ \bibnamefont
  {Silva}}, \bibinfo {author} {\bibfnamefont {W.~L.~C.}\ \bibnamefont {Lima}},
  \bibinfo {author} {\bibfnamefont {W.~C.}\ \bibnamefont {Oliveira}}, \bibinfo
  {author} {\bibfnamefont {J.~L.~N.}\ \bibnamefont {Mello}}, \bibinfo {author}
  {\bibfnamefont {L.~N.}\ \bibnamefont {Oliveira}}, \ and\ \bibinfo {author}
  {\bibfnamefont {J.~W.}\ \bibnamefont {Wilkins}},\ }\href {\doibase
  10.1103/PhysRevLett.76.275} {\bibfield  {journal} {\bibinfo  {journal} {Phys.
  Rev. Lett.}\ }\textbf {\bibinfo {volume} {76}},\ \bibinfo {pages} {275}
  (\bibinfo {year} {1996})}\BibitemShut {NoStop}%
\bibitem [{\citenamefont {Lechtenberg}\ \emph {et~al.}(2017)\citenamefont
  {Lechtenberg}, \citenamefont {Eickhoff},\ and\ \citenamefont
  {Anders}}]{lit:lechtenbergAnders17}%
  \BibitemOpen
  \bibfield  {author} {\bibinfo {author} {\bibfnamefont {B.}~\bibnamefont
  {Lechtenberg}}, \bibinfo {author} {\bibfnamefont {F.}~\bibnamefont
  {Eickhoff}}, \ and\ \bibinfo {author} {\bibfnamefont {F.~B.}\ \bibnamefont
  {Anders}},\ }\href {\doibase 10.1103/PhysRevB.96.041109} {\bibfield
  {journal} {\bibinfo  {journal} {Phys. Rev. B}\ }\textbf {\bibinfo {volume}
  {96}},\ \bibinfo {pages} {041109} (\bibinfo {year} {2017})}\BibitemShut
  {NoStop}%
\bibitem [{\citenamefont {Wilson}(1975)}]{lit:WilsonNRG}%
  \BibitemOpen
  \bibfield  {author} {\bibinfo {author} {\bibfnamefont {K.~G.}\ \bibnamefont
  {Wilson}},\ }\href {\doibase 10.1103/RevModPhys.47.773} {\bibfield  {journal}
  {\bibinfo  {journal} {Rev. Mod. Phys.}\ }\textbf {\bibinfo {volume} {47}},\
  \bibinfo {pages} {773} (\bibinfo {year} {1975})}\BibitemShut {NoStop}%
\bibitem [{\citenamefont {Bulla}\ \emph {et~al.}(2008)\citenamefont {Bulla},
  \citenamefont {Costi},\ and\ \citenamefont {Pruschke}}]{lit:BullaReview}%
  \BibitemOpen
  \bibfield  {author} {\bibinfo {author} {\bibfnamefont {R.}~\bibnamefont
  {Bulla}}, \bibinfo {author} {\bibfnamefont {T.~A.}\ \bibnamefont {Costi}}, \
  and\ \bibinfo {author} {\bibfnamefont {T.}~\bibnamefont {Pruschke}},\ }\href
  {\doibase 10.1103/RevModPhys.80.395} {\bibfield  {journal} {\bibinfo
  {journal} {Rev. Mod. Phys.}\ }\textbf {\bibinfo {volume} {80}},\ \bibinfo
  {pages} {395} (\bibinfo {year} {2008})}\BibitemShut {NoStop}%
\bibitem [{\citenamefont {Anders}\ and\ \citenamefont
  {Schiller}(2005)}]{lit:Anders05}%
  \BibitemOpen
  \bibfield  {author} {\bibinfo {author} {\bibfnamefont {F.~B.}\ \bibnamefont
  {Anders}}\ and\ \bibinfo {author} {\bibfnamefont {A.}~\bibnamefont
  {Schiller}},\ }\href {\doibase 10.1103/PhysRevLett.95.196801} {\bibfield
  {journal} {\bibinfo  {journal} {Phys. Rev. Lett.}\ }\textbf {\bibinfo
  {volume} {95}},\ \bibinfo {pages} {196801} (\bibinfo {year}
  {2005})}\BibitemShut {NoStop}%
\bibitem [{\citenamefont {Anders}\ and\ \citenamefont
  {Schiller}(2006)}]{lit:Anders06}%
  \BibitemOpen
  \bibfield  {author} {\bibinfo {author} {\bibfnamefont {F.~B.}\ \bibnamefont
  {Anders}}\ and\ \bibinfo {author} {\bibfnamefont {A.}~\bibnamefont
  {Schiller}},\ }\href {\doibase 10.1103/PhysRevB.74.245113} {\bibfield
  {journal} {\bibinfo  {journal} {Phys. Rev. B}\ }\textbf {\bibinfo {volume}
  {74}},\ \bibinfo {pages} {245113} (\bibinfo {year} {2006})}\BibitemShut
  {NoStop}%
\bibitem [{\citenamefont {Jones}\ and\ \citenamefont
  {Varma}(1987)}]{lit:Jones87}%
  \BibitemOpen
  \bibfield  {author} {\bibinfo {author} {\bibfnamefont {B.~A.}\ \bibnamefont
  {Jones}}\ and\ \bibinfo {author} {\bibfnamefont {C.~M.}\ \bibnamefont
  {Varma}},\ }\href {\doibase 10.1103/PhysRevLett.58.843} {\bibfield  {journal}
  {\bibinfo  {journal} {Phys. Rev. Lett.}\ }\textbf {\bibinfo {volume} {58}},\
  \bibinfo {pages} {843} (\bibinfo {year} {1987})}\BibitemShut {NoStop}%
\bibitem [{\citenamefont {Takasan}\ \emph {et~al.}(2017)\citenamefont
  {Takasan}, \citenamefont {Nakagawa},\ and\ \citenamefont
  {Kawakami}}]{lit:Takasan2017}%
  \BibitemOpen
  \bibfield  {author} {\bibinfo {author} {\bibfnamefont {K.}~\bibnamefont
  {Takasan}}, \bibinfo {author} {\bibfnamefont {M.}~\bibnamefont {Nakagawa}}, \
  and\ \bibinfo {author} {\bibfnamefont {N.}~\bibnamefont {Kawakami}},\ }\href
  {\doibase 10.1103/PhysRevB.96.115120} {\bibfield  {journal} {\bibinfo
  {journal} {Phys. Rev. B}\ }\textbf {\bibinfo {volume} {96}},\ \bibinfo
  {pages} {115120} (\bibinfo {year} {2017})}\BibitemShut {NoStop}%
\bibitem [{\citenamefont {Lieb}\ and\ \citenamefont
  {Robinson}(1972)}]{lit:LiebRobinsonBound72}%
  \BibitemOpen
  \bibfield  {author} {\bibinfo {author} {\bibfnamefont {E.~H.}\ \bibnamefont
  {Lieb}}\ and\ \bibinfo {author} {\bibfnamefont {D.~W.}\ \bibnamefont
  {Robinson}},\ }\href {\doibase 10.1007/BF01645779} {\bibfield  {journal}
  {\bibinfo  {journal} {Communications in Mathematical Physics}\ }\textbf
  {\bibinfo {volume} {28}},\ \bibinfo {pages} {251} (\bibinfo {year}
  {1972})}\BibitemShut {NoStop}%
\bibitem [{\citenamefont {Lechtenberg}\ and\ \citenamefont
  {Anders}(2014)}]{lit:lechtenbergAnders14}%
  \BibitemOpen
  \bibfield  {author} {\bibinfo {author} {\bibfnamefont {B.}~\bibnamefont
  {Lechtenberg}}\ and\ \bibinfo {author} {\bibfnamefont {F.~B.}\ \bibnamefont
  {Anders}},\ }\href {\doibase 10.1103/PhysRevB.90.045117} {\bibfield
  {journal} {\bibinfo  {journal} {Phys. Rev. B}\ }\textbf {\bibinfo {volume}
  {90}},\ \bibinfo {pages} {045117} (\bibinfo {year} {2014})}\BibitemShut
  {NoStop}%
\bibitem [{\citenamefont {Borda}(2007)}]{lit:Borda07}%
  \BibitemOpen
  \bibfield  {author} {\bibinfo {author} {\bibfnamefont {L.}~\bibnamefont
  {Borda}},\ }\href {\doibase 10.1103/PhysRevB.75.041307} {\bibfield  {journal}
  {\bibinfo  {journal} {Phys. Rev. B}\ }\textbf {\bibinfo {volume} {75}},\
  \bibinfo {pages} {041307} (\bibinfo {year} {2007})}\BibitemShut {NoStop}%
\bibitem [{\citenamefont {Jayaprakash}\ \emph {et~al.}(1981)\citenamefont
  {Jayaprakash}, \citenamefont {Krishna-murthy},\ and\ \citenamefont
  {Wilkins}}]{lit:JayaprakashKrischnamurtyWilkins1981}%
  \BibitemOpen
  \bibfield  {author} {\bibinfo {author} {\bibfnamefont {C.}~\bibnamefont
  {Jayaprakash}}, \bibinfo {author} {\bibfnamefont {H.~R.}\ \bibnamefont
  {Krishna-murthy}}, \ and\ \bibinfo {author} {\bibfnamefont {J.~W.}\
  \bibnamefont {Wilkins}},\ }\href {\doibase 10.1103/PhysRevLett.47.737}
  {\bibfield  {journal} {\bibinfo  {journal} {Phys. Rev. Lett.}\ }\textbf
  {\bibinfo {volume} {47}},\ \bibinfo {pages} {737} (\bibinfo {year}
  {1981})}\BibitemShut {NoStop}%
\bibitem [{\citenamefont {Eickhoff}\ \emph {et~al.}()\citenamefont {Eickhoff},
  \citenamefont {Lechtenberg},\ and\ \citenamefont
  {Anders}}]{lit:Eickhoff2018}%
  \BibitemOpen
  \bibfield  {author} {\bibinfo {author} {\bibfnamefont {F.}~\bibnamefont
  {Eickhoff}}, \bibinfo {author} {\bibfnamefont {B.}~\bibnamefont
  {Lechtenberg}}, \ and\ \bibinfo {author} {\bibfnamefont {F.~B.}\ \bibnamefont
  {Anders}},\ }\href@noop {} {\bibinfo  {journal} {arXiv:1806.03130}\
  }\BibitemShut {NoStop}%
\bibitem [{\citenamefont {Nghiem}\ and\ \citenamefont
  {Costi}(2014)}]{lit:Costi14}%
  \BibitemOpen
\bibfield  {journal} {  }\bibfield  {author} {\bibinfo {author} {\bibfnamefont
  {H.~T.~M.}\ \bibnamefont {Nghiem}}\ and\ \bibinfo {author} {\bibfnamefont
  {T.~A.}\ \bibnamefont {Costi}},\ }\href {\doibase 10.1103/PhysRevB.89.075118}
  {\bibfield  {journal} {\bibinfo  {journal} {Phys. Rev. B}\ }\textbf {\bibinfo
  {volume} {89}},\ \bibinfo {pages} {075118} (\bibinfo {year}
  {2014})}\BibitemShut {NoStop}%
\bibitem [{\citenamefont {Nghiem}\ and\ \citenamefont
  {Costi}(2017)}]{lit:Costi2017}%
  \BibitemOpen
  \bibfield  {author} {\bibinfo {author} {\bibfnamefont {H.~T.~M.}\
  \bibnamefont {Nghiem}}\ and\ \bibinfo {author} {\bibfnamefont {T.~A.}\
  \bibnamefont {Costi}},\ }\href {\doibase 10.1103/PhysRevLett.119.156601}
  {\bibfield  {journal} {\bibinfo  {journal} {Phys. Rev. Lett.}\ }\textbf
  {\bibinfo {volume} {119}},\ \bibinfo {pages} {156601} (\bibinfo {year}
  {2017})}\BibitemShut {NoStop}%
\bibitem [{\citenamefont {Anders}(2008)}]{lit:AndersSSnrg2008}%
  \BibitemOpen
  \bibfield  {author} {\bibinfo {author} {\bibfnamefont {F.~B.}\ \bibnamefont
  {Anders}},\ }\href {\doibase 10.1103/PhysRevLett.101.066804} {\bibfield
  {journal} {\bibinfo  {journal} {Phys. Rev. Lett.}\ }\textbf {\bibinfo
  {volume} {101}},\ \bibinfo {eid} {066804} (\bibinfo {year}
  {2008})}\BibitemShut {NoStop}%
\bibitem [{\citenamefont {Schmitt}\ and\ \citenamefont
  {Anders}(2010)}]{lit:SchmittAnders2009}%
  \BibitemOpen
  \bibfield  {author} {\bibinfo {author} {\bibfnamefont {S.}~\bibnamefont
  {Schmitt}}\ and\ \bibinfo {author} {\bibfnamefont {F.~B.}\ \bibnamefont
  {Anders}},\ }\href {\doibase 10.1103/PhysRevB.81.165106} {\bibfield
  {journal} {\bibinfo  {journal} {Phys. Rev. B}\ }\textbf {\bibinfo {volume}
  {81}},\ \bibinfo {pages} {165106} (\bibinfo {year} {2010})}\BibitemShut
  {NoStop}%
\bibitem [{\citenamefont {Schmitt}\ and\ \citenamefont
  {Anders}(2011)}]{lit:SchmittAnders2011}%
  \BibitemOpen
  \bibfield  {author} {\bibinfo {author} {\bibfnamefont {S.}~\bibnamefont
  {Schmitt}}\ and\ \bibinfo {author} {\bibfnamefont {F.~B.}\ \bibnamefont
  {Anders}},\ }\href@noop {} {\bibfield  {journal} {\bibinfo  {journal} {Phys.
  Rev. Lett.}\ }\textbf {\bibinfo {volume} {107}},\ \bibinfo {pages} {056801}
  (\bibinfo {year} {2011})}\BibitemShut {NoStop}%
\bibitem [{\citenamefont {Jovchev}\ and\ \citenamefont
  {Anders}(2013)}]{lit:JovchevAnders2013}%
  \BibitemOpen
  \bibfield  {author} {\bibinfo {author} {\bibfnamefont {A.}~\bibnamefont
  {Jovchev}}\ and\ \bibinfo {author} {\bibfnamefont {F.~B.}\ \bibnamefont
  {Anders}},\ }\href {\doibase 10.1103/PhysRevB.87.195112} {\bibfield
  {journal} {\bibinfo  {journal} {Phys. Rev. B}\ }\textbf {\bibinfo {volume}
  {87}},\ \bibinfo {pages} {195112} (\bibinfo {year} {2013})}\BibitemShut
  {NoStop}%
\bibitem [{\citenamefont {Eidelstein}\ \emph {et~al.}(2012)\citenamefont
  {Eidelstein}, \citenamefont {Schiller}, \citenamefont {G\"uttge},\ and\
  \citenamefont {Anders}}]{lit:Eidelstein2012}%
  \BibitemOpen
  \bibfield  {author} {\bibinfo {author} {\bibfnamefont {E.}~\bibnamefont
  {Eidelstein}}, \bibinfo {author} {\bibfnamefont {A.}~\bibnamefont
  {Schiller}}, \bibinfo {author} {\bibfnamefont {F.}~\bibnamefont {G\"uttge}},
  \ and\ \bibinfo {author} {\bibfnamefont {F.~B.}\ \bibnamefont {Anders}},\
  }\href {\doibase 10.1103/PhysRevB.85.075118} {\bibfield  {journal} {\bibinfo
  {journal} {Phys. Rev. B}\ }\textbf {\bibinfo {volume} {85}},\ \bibinfo
  {pages} {075118} (\bibinfo {year} {2012})}\BibitemShut {NoStop}%
\bibitem [{\citenamefont {G\"uttge}\ \emph {et~al.}(2013)\citenamefont
  {G\"uttge}, \citenamefont {Anders}, \citenamefont {Schollw\"ock},
  \citenamefont {Eidelstein},\ and\ \citenamefont
  {Schiller}}]{lit:Guettge2013}%
  \BibitemOpen
  \bibfield  {author} {\bibinfo {author} {\bibfnamefont {F.}~\bibnamefont
  {G\"uttge}}, \bibinfo {author} {\bibfnamefont {F.~B.}\ \bibnamefont
  {Anders}}, \bibinfo {author} {\bibfnamefont {U.}~\bibnamefont
  {Schollw\"ock}}, \bibinfo {author} {\bibfnamefont {E.}~\bibnamefont
  {Eidelstein}}, \ and\ \bibinfo {author} {\bibfnamefont {A.}~\bibnamefont
  {Schiller}},\ }\href {\doibase 10.1103/PhysRevB.87.115115} {\bibfield
  {journal} {\bibinfo  {journal} {Phys. Rev. B}\ }\textbf {\bibinfo {volume}
  {87}},\ \bibinfo {pages} {115115} (\bibinfo {year} {2013})}\BibitemShut
  {NoStop}%
\bibitem [{\citenamefont {Yoshida}\ \emph {et~al.}(1990)\citenamefont
  {Yoshida}, \citenamefont {Whitaker},\ and\ \citenamefont
  {Oliveira}}]{lit:Yoshida90}%
  \BibitemOpen
  \bibfield  {author} {\bibinfo {author} {\bibfnamefont {M.}~\bibnamefont
  {Yoshida}}, \bibinfo {author} {\bibfnamefont {M.~A.}\ \bibnamefont
  {Whitaker}}, \ and\ \bibinfo {author} {\bibfnamefont {L.~N.}\ \bibnamefont
  {Oliveira}},\ }\href {\doibase 10.1103/PhysRevB.41.9403} {\bibfield
  {journal} {\bibinfo  {journal} {Phys. Rev. B}\ }\textbf {\bibinfo {volume}
  {41}},\ \bibinfo {pages} {9403} (\bibinfo {year} {1990})}\BibitemShut
  {NoStop}%
\bibitem [{\citenamefont {Allerdt}\ \emph {et~al.}(2015)\citenamefont
  {Allerdt}, \citenamefont {B\"usser}, \citenamefont {Martins},\ and\
  \citenamefont {Feiguin}}]{lit:Feiguin2015}%
  \BibitemOpen
  \bibfield  {author} {\bibinfo {author} {\bibfnamefont {A.}~\bibnamefont
  {Allerdt}}, \bibinfo {author} {\bibfnamefont {C.~A.}\ \bibnamefont
  {B\"usser}}, \bibinfo {author} {\bibfnamefont {G.~B.}\ \bibnamefont
  {Martins}}, \ and\ \bibinfo {author} {\bibfnamefont {A.~E.}\ \bibnamefont
  {Feiguin}},\ }\href {\doibase 10.1103/PhysRevB.91.085101} {\bibfield
  {journal} {\bibinfo  {journal} {Phys. Rev. B}\ }\textbf {\bibinfo {volume}
  {91}},\ \bibinfo {pages} {085101} (\bibinfo {year} {2015})}\BibitemShut
  {NoStop}%
\bibitem [{\citenamefont {{Zhu}}\ and\ \citenamefont
  {{Varma}}(2006)}]{lit:Varma2007}%
  \BibitemOpen
  \bibfield  {author} {\bibinfo {author} {\bibfnamefont {L.}~\bibnamefont
  {{Zhu}}}\ and\ \bibinfo {author} {\bibfnamefont {C.~M.}\ \bibnamefont
  {{Varma}}},\ }\href@noop {} {\bibfield  {journal} {\bibinfo  {journal}
  {eprint arXiv:cond-mat/0607426}\ } (\bibinfo {year} {2006})},\ \Eprint
  {http://arxiv.org/abs/cond-mat/0607426} {cond-mat/0607426} \BibitemShut
  {NoStop}%
\bibitem [{\citenamefont {Barzykin}\ and\ \citenamefont
  {Affleck}(1998)}]{lit:Barzykin1998}%
  \BibitemOpen
  \bibfield  {author} {\bibinfo {author} {\bibfnamefont {V.}~\bibnamefont
  {Barzykin}}\ and\ \bibinfo {author} {\bibfnamefont {I.}~\bibnamefont
  {Affleck}},\ }\href {\doibase 10.1103/PhysRevB.57.432} {\bibfield  {journal}
  {\bibinfo  {journal} {Phys. Rev. B}\ }\textbf {\bibinfo {volume} {57}},\
  \bibinfo {pages} {432} (\bibinfo {year} {1998})}\BibitemShut {NoStop}%
\bibitem [{\citenamefont {Ishii}(1978)}]{lit:Ishii1978}%
  \BibitemOpen
  \bibfield  {author} {\bibinfo {author} {\bibfnamefont {H.}~\bibnamefont
  {Ishii}},\ }\href {\doibase 10.1007/BF00117963} {\bibfield  {journal}
  {\bibinfo  {journal} {Journal of Low Temperature Physics}\ }\textbf {\bibinfo
  {volume} {32}},\ \bibinfo {pages} {457} (\bibinfo {year} {1978})}\BibitemShut
  {NoStop}%
\bibitem [{\citenamefont {Anderson}(1970)}]{lit:PWAnderson1970_II}%
  \BibitemOpen
  \bibfield  {author} {\bibinfo {author} {\bibfnamefont {P.~W.}\ \bibnamefont
  {Anderson}},\ }\href {http://stacks.iop.org/0022-3719/3/i=12/a=008}
  {\bibfield  {journal} {\bibinfo  {journal} {Journal of Physics C: Solid State
  Physics}\ }\textbf {\bibinfo {volume} {3}},\ \bibinfo {pages} {2436}
  (\bibinfo {year} {1970})}\BibitemShut {NoStop}%
\bibitem [{\citenamefont {Nejati}\ \emph {et~al.}(2017)\citenamefont {Nejati},
  \citenamefont {Ballmann},\ and\ \citenamefont {Kroha}}]{lit:Kroha17}%
  \BibitemOpen
  \bibfield  {author} {\bibinfo {author} {\bibfnamefont {A.}~\bibnamefont
  {Nejati}}, \bibinfo {author} {\bibfnamefont {K.}~\bibnamefont {Ballmann}}, \
  and\ \bibinfo {author} {\bibfnamefont {J.}~\bibnamefont {Kroha}},\ }\href
  {\doibase 10.1103/PhysRevLett.118.117204} {\bibfield  {journal} {\bibinfo
  {journal} {Phys. Rev. Lett.}\ }\textbf {\bibinfo {volume} {118}},\ \bibinfo
  {pages} {117204} (\bibinfo {year} {2017})}\BibitemShut {NoStop}%
\bibitem [{Note1()}]{Note1}%
  \BibitemOpen
  \bibinfo {note} {For large couplings $\rho J> 0.50$ a second minimum prior to
  the first one slowly starts to develop the value of which may even become
  smaller than the value of the original second minimum for very large
  couplings $\rho J>0.55$. In order to achieve comparability with the curves
  for smaller couplings, we thus use the value of the second minimum as
  $<\protect \mathaccentV {vec}17E{S}_1\protect \mathaccentV
  {vec}17E{S}_2>_{\protect \text {min}}$ for couplings $\rho
  J>0.55$.}\BibitemShut {Stop}%
\bibitem [{\citenamefont {Hanson}\ \emph {et~al.}(2007)\citenamefont {Hanson},
  \citenamefont {Kouwenhoven}, \citenamefont {Petta}, \citenamefont {Tarucha},\
  and\ \citenamefont {Vandersypen}}]{lit:Hanson2007}%
  \BibitemOpen
  \bibfield  {author} {\bibinfo {author} {\bibfnamefont {R.}~\bibnamefont
  {Hanson}}, \bibinfo {author} {\bibfnamefont {L.~P.}\ \bibnamefont
  {Kouwenhoven}}, \bibinfo {author} {\bibfnamefont {J.~R.}\ \bibnamefont
  {Petta}}, \bibinfo {author} {\bibfnamefont {S.}~\bibnamefont {Tarucha}}, \
  and\ \bibinfo {author} {\bibfnamefont {L.~M.~K.}\ \bibnamefont
  {Vandersypen}},\ }\href {\doibase 10.1103/RevModPhys.79.1217} {\bibfield
  {journal} {\bibinfo  {journal} {Rev. Mod. Phys.}\ }\textbf {\bibinfo {volume}
  {79}},\ \bibinfo {pages} {1217} (\bibinfo {year} {2007})}\BibitemShut
  {NoStop}%
\bibitem [{\citenamefont {Merkulov}\ \emph {et~al.}(2002)\citenamefont
  {Merkulov}, \citenamefont {Efros},\ and\ \citenamefont
  {Rosen}}]{lit:Merkulov2002}%
  \BibitemOpen
  \bibfield  {author} {\bibinfo {author} {\bibfnamefont {I.~A.}\ \bibnamefont
  {Merkulov}}, \bibinfo {author} {\bibfnamefont {A.~L.}\ \bibnamefont {Efros}},
  \ and\ \bibinfo {author} {\bibfnamefont {M.}~\bibnamefont {Rosen}},\ }\href
  {\doibase 10.1103/PhysRevB.65.205309} {\bibfield  {journal} {\bibinfo
  {journal} {Phys. Rev. B}\ }\textbf {\bibinfo {volume} {65}},\ \bibinfo
  {pages} {205309} (\bibinfo {year} {2002})}\BibitemShut {NoStop}%
\bibitem [{\citenamefont {Coish}\ and\ \citenamefont
  {Loss}(2004)}]{lit:Loss2004}%
  \BibitemOpen
  \bibfield  {author} {\bibinfo {author} {\bibfnamefont {W.~A.}\ \bibnamefont
  {Coish}}\ and\ \bibinfo {author} {\bibfnamefont {D.}~\bibnamefont {Loss}},\
  }\href {\doibase 10.1103/PhysRevB.70.195340} {\bibfield  {journal} {\bibinfo
  {journal} {Phys. Rev. B}\ }\textbf {\bibinfo {volume} {70}},\ \bibinfo
  {pages} {195340} (\bibinfo {year} {2004})}\BibitemShut {NoStop}%
\bibitem [{\citenamefont {Fischer}\ \emph {et~al.}(2008)\citenamefont
  {Fischer}, \citenamefont {Coish}, \citenamefont {Bulaev},\ and\ \citenamefont
  {Loss}}]{lit:Loss2008}%
  \BibitemOpen
  \bibfield  {author} {\bibinfo {author} {\bibfnamefont {J.}~\bibnamefont
  {Fischer}}, \bibinfo {author} {\bibfnamefont {W.~A.}\ \bibnamefont {Coish}},
  \bibinfo {author} {\bibfnamefont {D.~V.}\ \bibnamefont {Bulaev}}, \ and\
  \bibinfo {author} {\bibfnamefont {D.}~\bibnamefont {Loss}},\ }\href {\doibase
  10.1103/PhysRevB.78.155329} {\bibfield  {journal} {\bibinfo  {journal} {Phys.
  Rev. B}\ }\textbf {\bibinfo {volume} {78}},\ \bibinfo {pages} {155329}
  (\bibinfo {year} {2008})}\BibitemShut {NoStop}%
\end{thebibliography}
%

\end{document}